\documentclass[article,onefignum,onetabnum]{siamart171218}
\usepackage{todonotes}

\usepackage{lipsum}
\usepackage{amsfonts}
\usepackage{graphicx}
\usepackage{cite}
\usepackage{epstopdf}
\usepackage{algorithmic}
\usepackage[caption=false]{subfig}
\ifpdf
  \DeclareGraphicsExtensions{.eps,.pdf,.png,.jpg}
\else
  \DeclareGraphicsExtensions{.eps}
\fi

\newsiamremark{remark}{Remark}
\newsiamremark{hypothesis}{Hypothesis}
\crefname{hypothesis}{Hypothesis}{Hypotheses}
\newsiamthm{claim}{Claim}

\headers{Spatiotemporal Hawkes Processes and Network Reconstruction}{B. Yuan, H. Li, A. L. Bertozzi, P. J. Brantingham and M. A. Porter}

\title{Multivariate Spatiotemporal Hawkes Processes and Network Reconstruction\thanks{Submitted to the editors DATE.
\funding{This work is supported by NSF grant DMS-1417674, NSF grant DMS-1737770, and DARPA grant FA8750-18-2-0066.}}}

\author{
	Baichuan Yuan\thanks{Department of Mathematics, University of California, Los Angeles, Los Angeles, CA
		(\email{byuan@math.ucla.edu}, \email{lihao0809@math.ucla.edu}, \email{bertozzi@math.ucla.edu}, \email{mason@math.ucla.edu}).}
	\and
	Hao Li\footnotemark[2]
	\and
	Andrea L. Bertozzi\footnotemark[2]
	\and
	P. Jeffrey Brantingham\thanks{Department of Anthropology, University of California, Los Angeles, Los Angeles, CA
		(\email{branting@ucla.edu}).}
	\and
	Mason A. Porter\footnotemark[2]
}

\usepackage{amsopn}

\ifpdf
\hypersetup{
  pdftitle={MULTIVARIATE SPATIOTEMPORAL HAWKES PROCESSES AND NETWORK RECONSTRUCTION},
  pdfauthor={B. Yuan, H. Li, A. L. Bertozzi, P. J. Brantingham and M. A. Porter}
}
\fi

\begin{document}

\maketitle

\begin{abstract}
There is often latent network structure in spatial and temporal data and the tools of network analysis can yield fascinating insights into such data. In this paper, we develop a nonparametric method for network reconstruction from spatiotemporal data sets using multivariate Hawkes processes. In contrast to prior work on network reconstruction with point-process models, which has often focused on exclusively temporal information, our approach uses both temporal and spatial information and does not assume a specific parametric form of network dynamics. This leads to an effective way of recovering an underlying network. We illustrate our approach using both synthetic networks and networks constructed from real-world data sets (a location-based social media network, a narrative of crime events, and violent gang crimes).
Our results demonstrate that, in comparison to using only temporal data, our spatiotemporal approach yields improved network reconstruction, providing a basis for meaningful subsequent analysis --- such as community structure and motif analysis --- of the reconstructed networks.
\end{abstract}





\section{Introduction} \label{intro}

Digital devices such as smartphones and tablets generate a massive amount of spatiotemporal data about human activities, providing a wonderful opportunity for researchers to gain insight into human dynamics through our ``digital footprints''. A broad variety of human activities are analyzed using such data, creating new disciplines \cite{Lazer721} such as computational social science and digital humanities. Examples of such activities include online check-ins in large cities \cite{cho_friendship_2011}, effects of human mobility \cite{balcan2009multiscale} and currency flow \cite{brockmann2006} on the spread of contagious diseases, online communications during Occupy Wall Street\cite{conover2013geospatial}, crime reports in Los Angeles county \cite{kuang_crime_2017}, and many others.

Network analysis is a powerful approach for representing and analyzing complex systems of interacting components \cite{newman2018}, and network-based methods can provide considerable insights into the structure and dynamics of complex spatiotemporal data \cite{barthelemy2018}. It has been valuable for studies of both digital human footprints and human mobility \cite{barbosa2018}. To give one recent example, Noulas et al. \cite{noulas2012tale} studied geographic online social networks to illustrate similarities and heterogeneities in human mobility patterns.

Suppose that each node in a network represents an entity, and that the edges (which can be either undirected or directed, and can be either unweighted or weighted) represent spatiotemporal connections between pairs of entities. For instance, in a check-in data set from a social medium, one can model each user as a node, which has associated check-in time and locations. In this case, one can suppose that an edge exists between a pair of users if they follow each other on the social medium. One can use edge weights to quantify the amount of ``influence'' between users, where a larger weight signifies a larger impact. In our investigation, we assume that the relationships between nodes are time-independent.\footnote{For other regimes of relative time scales between spatiotemporal processes and network dynamics, it is necessary to consider time-dependent edges \cite{porter2016dynamical,holme2015}.}
In some cases, the entities and relationships are both known, and one can investigate the structure and dynamics of the associated networks. However, in many situations, network data is incomplete --- with potentially a large amount of missing data, in the form of missing entities, interactions, and/or metadata \cite{stomakhin2011reconstruction} --- and the relationships between nodes may not be directly observable \cite{scellato2011exploiting}. For example, social-media companies attempt to infer friendship relationships among their users to provide accurate friendship recommendations for online social networks. 

In the last few years, there has been a considerable amount of work on inferring missing data (both structure and weights) in networks.
A basic approach for inferring relationships among entities is to calculate cross-correlations of their associated time series \cite{lauritzen1996graphical}. Another approach is to use coefficients from a generalized linear model (GLM)\cite{nelder1972generalized}, a generalization of linear regression that allows response variables to have a non-Gaussian error-distribution. 
Recently, people have begun to use point-process methods \cite{simma2010modeling} in network reconstruction. 
For example, Perry and Wolfe \cite{perry_point_2013} modeled networks as a multivariate point process and then inferred covariate-based edges (both their existence and their weights) by estimating a point process. Among point-process models, it is very popular to use Hawkes processes (also known as self-exciting point processes\footnote{We use these terms interchangeably in this paper.} ) for studying human dynamics \cite{linderman_discovering_2014,fox_modeling_2016}. Hawkes-process models are characterized by mutual ``triggering'' among events \cite{ogata_statistical_1988}, as one event may increase the probability for subsequent events to occur. Such models can capture inhomogeneous inter-event times and causal (temporal) correlations, which have both been observed in human dynamics \cite{kivela2015estimating}. These properties make it a useful approach in social-network applications \cite{karsai2018bursty}.
It thus seems promising to use such processes for network inference on dynamic human data, such as crime events or online social activity. For example, Linderman and Adams \cite{linderman_discovering_2014} proposed a fully-bayesian Hawkes model that they reported to be more accurate for their data at inferring missing edges than GLMs, cross-correlations, and a simple self-exciting point process with an exponential kernel. Very recently, self-exciting point processes were applied in \cite{suny2018inferring} to reconstruct multilayer networks \cite{kivela2014multilayer}, a generalization of ordinary graphs. However, the aforementioned temporal point-process models are not without limitations.
For example, most of these models do not use spatial information, even when it plays a significant role in a system's dynamics. Furthermore, many assume an a priori model\cite{linderman_discovering_2014} or a specific parametrization \cite{tita_latent_2014} for their point processes.

In the present paper, we propose a nonparametric and multivariate version of the spatiotemporal Hawkes process. Spatiotemporal Hawkes processes have been used previously to study numerous topics, including crime \cite{mohler_self-exciting_2011}, social media \cite{lai2016topic}, and earthquake prediction \cite{fox2016spatially}. In our model, each node in a network is associated with a spatiotemporal Hawkes process. The nodes can ``trigger'' each other, so events that are associated with one node increase the probability that there will be events associated with the other nodes. We measure the extent of such mutual-triggering effects using a $U \times U$ ``triggering matrix'' $\mathbf{K}$, where $U$ is the number of nodes. If one considers an exclusively temporal scenario, a point process $u$ does not ``cause'' (in the Granger sense \cite{granger1969investigating}) a point process $v$ if and only if $\mathbf{K}(u,v)=0$ \cite{eichler2017graphical}. Because triggering between point processes reflects an underlying connection, one can try to recover latent relationships in a network from $\mathbf{K}$. Such triggering should decrease with both distance and time according to some spatial and temporal kernels. In our work, instead of assuming exponential decay \cite{fox_modeling_2016} or some other distribution \cite{linderman_discovering_2014, tita_latent_2014}, we adopt a nonparametric approach \cite{marsan2008extending} to learn both spatial and temporal kernels from data using an expectation-maximization-type (EM-type) algorithm \cite{veen2008estimation}. 

We compare our approach with other recent point-process network-reconstruction methods \cite{fox_modeling_2016,linderman_discovering_2014} on both synthetic and real-world data sets with spatial information. Our two examples of the latter data sets come from a location-based social-networking website and crime topics.
 We illustrate the importance both of incorporating spatial information and of using nonparametric kernels. Although we assume that the relationships among nodes are time-independent, our model still recovers a causal structure among events in synthetic data sets. We also build event-causality networks on data sets about violent crimes of gangs and examine gang retaliation patterns using motif analysis. 

Our paper proceeds as follows. In \cref{sec:review}, we review self-exciting point processes and recent point-process methods for network reconstruction. In \cref{sec:alg}, we introduce our nonparametric spatiotemporal model and our approaches to model estimation and simulation. In \cref{sec:experiments}, we compare our model with others on both synthetic and real-world data sets. We conclude in \cref{sec:conclusions}. We give details about our preprocessing for the social-networking website data set in Appendix \ref{appA}.

\section{Self-Exciting Point Processes}
\label{sec:review}

A \textit{point process} $S$ is a random measure on a complete separable metric space that takes values on $\{0,1, 2, \ldots \} \cup \{ \infty\}$  \cite{schoenberg2002point}. We first consider a \textit{temporal point process}, which consists of a list $\{t_1,t_2,\ldots,t_N\}$ of $N$ time points, with corresponding events $1,2,\ldots,N$. Let $S[a,b)$ denote the number of points (i.e., events) that occur in a finite time interval $[a,b)$, with $a<b$. One typically models the behavior of a simple temporal point process (multiple events cannot occur at the same time) by specifying its conditional intensity function $\lambda(t)$, which represents the rate at which events are expected to occur around a particular time $t$, conditional on the prior history of the point process before time $t$. Specifically, when $H_t = \{ t_i|t_i<t \}$ is the history of the process up to time $t$, one defines the \textit{conditional intensity function}
\begin{equation*}
    \lambda(t) = \lim_{\Delta t \downarrow 0} \frac{\mathbb{E}[S[t,t+\Delta t)|H_t]}{\Delta t}\,.
\end{equation*}
One important point-process model is a \textit{Poisson process}, in which the number of points in any time interval follows a Poisson distribution and the number of points in disjoint sets are independent. A Poisson process is called \textit{homogeneous} if $\lambda(t) \equiv \mathrm{constant}$ and is thus characterized by a constant rate at which events are expected to occur per unit time. It is called \textit{inhomogeneous} if the conditional intensity function $\lambda(t)$ depends on the time $t$ (e.g., $\lambda(t)=\sin(t)$). In both situations, the numbers of points (i.e., events) in disjoint intervals are independent random variables.
 
We now discuss self-exciting point processes, which allow one to examine a notion of causality in a point process.  
If we consider a list $\{t_1,t_2,\ldots,t_N\}$ of time stamps, we say that a point process is \textit{self-exciting} if
\begin{equation*}
	\text{Cov}\left[S(t_{k-1},t_{k}),S(t_{k},t_{k+1})\right] > 0 \quad \text{for}\;\; k \;\; \text{such that} \quad t_{k-1}<t_k<t_{k+1}\,.
\end{equation*}
That is, if an event occurs, another event becomes more likely to occur locally in time. 

A \emph{univariate temporal Hawkes process} has the following conditional intensity function:
\begin{equation}\label{thesum}
	\lambda(t)= \mu(t)+K \sum_{t_k<t} g(t-t_k)\,,
\end{equation}
where the background rate $\mu(t)$ can either be a constant or a time-dependent function that describes how the likelihood of some process (crimes, e-mails, tweets, and so on) evolves in time. For example, violent crimes are more likely to happen at night than during the day, and business e-mails are less likely to be sent during the weekend than on a weekday. One can construe the rate $\mu(t)$ as a process that designates the likelihood of an event to occur, independent of the other events. The summation term in Equation \cref{thesum} describes the self-excitation: past events increase the current conditional intensity. The function $g(t)$ is called the \emph{triggering kernel}, and the parameter $K$ denotes the mean number of events that are triggered by an event. One standard example is a Hawkes process with an exponential kernel $g(t)=\omega e^{- \omega t}$, where $\omega$ is a constant decay rate for the triggering kernel that controls how fast the rate $\lambda(t)$ returns to its baseline level $\mu(t)$ after an event occurs. 


\subsection{Temporal Multivariate Models}

In network reconstruction, one seeks to infer the relationships (i.e., edges) and the strengths of such relationships (i.e., edge weights) among a set of entities (i.e., nodes). When modeling the relationships in a network, it is more appropriate to use a multivariate point process than a univariate one. In a temporal multivariate point process, there are $U$ different point processes $(S_u)_{u=1,\ldots,U}$, and the corresponding conditional intensity functions are $(\lambda_u(t))_{u=1,\ldots,U}$. We seek to infer the intensity functions from observed data $(t_j, u_j)_{j=1,\ldots,N}$ in a time window $[0,T]$, where $t_j$ and $u_j$, respectively, are the time and point-process index of event $j$. There are numerous applications of temporal multivariate point processes, such as financial markets \cite{bacry_hawkes_2015}, real-time crime forecasting\cite{wang2018graph} and neural spike trains \cite{brown2004multiple}. Here we focus on the specific application of network reconstruction.
  
A trivial example of a multivariate point process is the multivariate Poisson process, in which each point process is a univariate Poisson process. Another example is the multivariate Cox process, which consists of doubly stochastic Poisson processes in which the conditional intensity itself is a stochastic process. Perry and Wolfe \cite{perry_point_2013} used a Cox process to model e-mail interactions (edges) among a set of users (nodes). Neither the multivariate Poisson nor the multivariate Cox process are self-exciting.

Instead of modeling edges as Cox processes, Fox et al. \cite{fox_modeling_2016} used multivariate Hawkes processes to model people (nodes) communicating with each other via e-mail. Their conditional intensity function has an exponential kernel and a nonparametric background function $\mu_u(t)$ for each person (process) $u$:
\begin{equation} \label{eq:multi}
	\lambda_u(t)=\mu_u(t)+\sum_{t_i<t} K_{u_iu} \omega e^{-\omega(t-t_i)}\,,
\end{equation}	
where $K_{uv}=\mathbf{K}(u,v)$ is the expected number of events of person $v$ that are triggered by one event of person $u$. One can estimate the set of parameters $\Theta$ by minimizing the negative log-likelihood function 
\begin{equation}
    \label{eq: nmle}
	-\log(L(\Theta)) = -\sum_{k=1}^{N} \log(\lambda_{u_k}(t_k)) + \sum_{u=1}^{U}\int_0^T \lambda_u(t) \mathrm{d}t\,.
\end{equation}
Recall that $u_k$ is the point process associated with event $k$.

There are several variants of the multivariate Hawkes process. One is to add regularization terms to Equation \cref{eq: nmle} to improve the accuracy of parameter estimation. Lewis and Mohler \cite{lewis2011nonparametric} used maximum-penalized likelihood estimation, which enforces some regularity on the model parameters, to infer Hawkes processes. Zhou et al. \cite{zhou_learning_2013-1} extended this idea and promoted the low-rank and sparsity properties of $\mathbf{K}$ by adding nuclear and $L_1$ norms of $\mathbf{K}$ to Equation \cref{eq: nmle} with the conditional intensity function $\lambda_u(t)$ from Equation \cref{eq:multi}. Linderman et al. \cite{linderman_discovering_2014}
added random-graph priors on $\mathbf{K}$ and developed a fully Bayesian multivariate Hawkes model. See \cite{mark2018network} for theoretical guarantees on inferring Hawkes processes with a regularizer. Another research direction is to speed up the parameter estimation of point-process models. For example, Hall et al. \cite{hall2016tracking} tried to learn the triggering matrix $\mathbf{K}$ via an online learning framework for streaming data. Instead of using a likelihood-based method, Achab et al. \cite{achab2018uncovering} developed a fast moment-matching method to estimate the matrix $\mathbf{K}$.


\subsection{Spatiotemporal Point Processes}

Many real-world data sets include not only time stamps but also accompanying spatial information, which can be particularly important for correctly inferring and understanding the associated dynamics \cite{barthelemy2018}. In earthquakes, for example, most aftershocks usually occur geographically near the main shock \cite{ogata_space-time_1998}. In online social media, if two people often check in at the same location at closely proximate times, there is more likely to be a connection between them than if such ``joint check-ins'' occur rarely \cite{cho_friendship_2011}. These situations suggest that it is important to examine spatiotemporal point processes, rather than just temporal ones. Indeed, there are myriad applications of spatiotemporal Hawkes processes, including crime prediction \cite{mohler_self-exciting_2011}, seismology  \cite{ogata_space-time_1998}, and Twitter topics \cite{lai2016topic}. The successful employment of such processes in earthquake prediction and predictive policing \cite{mohler2015randomized} have helped inspire our work, in which we extend these ideas to network reconstruction.

We characterize a spatiotemporal point process $S(t,x,y)$ via its conditional intensity $\lambda(t,x,y)$, which is the expected rate of the accumulation of points around a particular spatiotemporal location. Given the history $\mathcal{H}_t$ of all points up to time $t$, we write
\begin{equation*}
	\lambda(t,x,y) = \lim_{\Delta t, \Delta x, \Delta y \downarrow 0} \left( \dfrac{\mathbb{E}\left[S\{(t,t+\Delta t) \times (x,x+\Delta x) \times (y,y+\Delta y )\}| \mathcal{H}_t \right]}{\Delta t \, \Delta x \, \Delta y} \right)\,.
\end{equation*}
For the purpose of modeling earthquakes, \cite{ogata_space-time_1998} used a self-exciting point process with a conditional intensity of the form 
\begin{equation*}
	\lambda(t,x,y) = \mu(x,y) + \sum_{t>t_i} g(x-x_i,y-y_i,t-t_i)\,.
\end{equation*}
In this setting, if an earthquake occurs, aftershocks are more likely to occur locally in time and space. The choice of the triggering kernel $g(t,x,y)$ is inspired by physical properties of earthquakes. For example, \cite{ogata_space-time_1998} used a modified Omori formula (a power law) \cite{ogata_statistical_1988} to describe the frequency of aftershocks per unit time.
In sociological applications, there is no direct theory to indicate appropriate choices for the kernel function. Some researchers have chosen specific kernels (e.g., exponential kernels) that are easy to compute. For example, Tita et al.~\cite{tita_latent_2014} used a spatiotemporal point process to infer missing information about event participants. They modeled interactions between event participants as a combination of a spatial Gaussian mixture model and a temporal Hawkes process with an exponential kernel. A key problem is how to justify kernel choices in specific applications.

\section{Spatiotemporal Models for Network Reconstruction} \label{sec:alg}

Many network-reconstruction methods using self-exciting point processes, such as \cite{linderman_discovering_2014,fox_modeling_2016}, have inferred time-independent relationships (i.e., edges) among entities (i.e., nodes) with corresponding (exclusively) temporal point processes. 
Entity (process) $u$ is adjacent to $v$ if $\mathbf{K}(u,v)>0$, where one estimates the triggering matrix $\mathbf{K}$ from the data. Entity $u$ is not adjacent to $v$ if entity $u$'s point process does not cause entity $v$'s point process in time (in the Granger sense \cite{eichler2017graphical}). For many problems, it is desirable --- or even crucial --- to incorporate spatial information \cite{cressie2015statistics,barthelemy2018}. For example, spatial information is an important part of online fingerprints in human activity, and it has a significant impact on most other social networks. In crime modeling, for example, there is a ``near repeat'' phenomenon in crime locations, indicating the necessity of including spatial information. Specifically, the spatial neighborhood of an initial burglary has a higher risk of repeat victimization than more-distant locations \cite{short2010dissipation}. In our work, we propose multivariate spatiotemporal Hawkes processes to infer relationships in networks and provide a novel approach for analyzing spatiotemporal dynamics.

Another important issue is the assumptions on triggering kernels for a Hawkes process. In seismology, for example, researchers attempt to use an underlying physical model to help determine a good kernel. However, it is much more difficult to validate such models in social networks than for physical or even biological phenomena \cite{porter2017role}. The content of social data is often unclear, and there is often little understanding of the underlying mechanisms that produce them. With less direct knowledge of possible triggering kernels, it is helpful to employ a data-driven method for kernel selection. Using a kernel with an inappropriate decay rate may lead to either underestimation or overestimation of the elements in the triggering matrix $\mathbf{K}$, which may also include false negatives or positives in the inferred relationships between entities.
Therefore, we ultimately use a nonparametric approach to learn triggering kernels in various applications to avoid a priori assumptions about a specific parametrization.

A multivariate spatiotemporal Hawkes process is a sequence 
$\{(t_i,x_i,y_i, u_i)\}_{i=1}^N$ with $N$ events, where $t_i$ and $(x_i,y_i)$ are spatiotemporal stamps and $u_i$ is the point-process index of event $i$. Each of the $U$ nodes is a \textit{marginal process}. The conditional intensity function for node $u$ is
\begin{equation}\label{spteH}
	\lambda_u(t,x,y) = \mu_u(x,y) + \sum_{t>t_i}  K_{u_iu} g(x-x_i,y-y_i,t-t_i)\,.
\end{equation}
The above Hawkes process assumes that each node $u$ has a background Poisson process that is constant in time but inhomogeneous in space with conditional intensity $\mu_u(x,y)$. There is also self-excitation, as past events increase the likelihood of subsequent events. We quantify the amount of impact that events associated with node $u_i$ have on subsequent events of node $u_j$ with a spatiotemporal kernel and the element $\textbf{K}(u_i,u_j)=K_{u_iu_j}$ of the triggering matrix. 


\subsection{A Parametric Model} \label{subsec: a parametric model}

We first propose a multivariate Hawkes process with a specific parametric form. We use this model to generate spatiotemporal events on synthetic networks and provide a form of ``ground truth'' that we can use later. 

The background rate $\mu_u$ and the triggering kernel $g$ for Equation \cref{spteH} are given by
\begin{align*}
	g(x,y,t) &= g_1(t) \times g_2(x,y) = \omega \exp\left(-\omega t\right) \times \dfrac{1}{2\pi \sigma^2} \exp \left(-\dfrac{x^2+y^2}{2 \sigma^2}\right)\,, \\
	\mu_u(x,y) &= \sum_{i=1}^N \dfrac{\beta_{u_iu}}{2 \pi \eta^2T}\times \exp\left(-\dfrac{(x-x_i)^2+(y-y_i)^2}{2 \eta^2}\right).
\end{align*}
For simplicity, we use exponential decay in time \cite{ogata_statistical_1988} and a Gaussian kernel in space \cite{mohler2014marked}. We let $T$ denote the time window of a data set; $K_{u_iu}$ denote the mean number of the events in process $u$ that are triggered by each event in the process $u_i$; the quantity $\beta_{u_iu}$ denote the extent to which events in process $u_i$ contribute to the background rate for events in the process $u$; and $\sigma$ and $\eta$, respectively, denote the standard deviations in the triggering kernel and background rate. The value of $\sigma$ determines the spreading scale of the triggering effect in space.  


\subsection{A Nonparametric Model}

With the conditional intensity given in Equation \cref{spteH}, we estimate the triggering kernel $g(x,y,t)=g_1(t)\times g_2(x,y)$ nonparametrically  using histogram estimators\cite{marsan2008extending}. We assume that $g_2$ is isotropic, which entails that $g_2(x,y) = g_2(r)$, where $r = \sqrt{x^2 + y^2}$. We let $h(r)$ be the spatial triggering kernel in the polar coordinate: $h(r)=2\pi r g_2(r)$. We extend the background rate that was proposed in \cite{fox2016spatially} to the multivariate case and write
\begin{equation}
	\mu_u(x,y)=\gamma_u \tau(x,y)=\frac{\gamma_u}{T} \sum_{i=1}^N \frac{p_{ii} }{2\pi d_i^2}\exp\left(-\frac{(x-x_i)^2+(y-y_i)^2} {2d_i^2}\right)\, ,\label{eq:nonpara_cond}
\end{equation}
where $\gamma_u$ is the background intensity of process $u$ and $p_{ii}$ is the probability that event $i$ is a background event (i.e., it is not triggered by any event). We compute $d_i$ by determining the radius of the smallest disk centered at $(x_i, y_i)$ that includes at least $n_p$ other events and is at least as large as some small value $\epsilon$ that represents the error in location. 

Once we fit the model to spatiotemporal data, the triggering matrix $\mathbf{K}$ gives our inferences for the underlying relationships between entities. For two entities $u$ and $v$, the matrix element $\mathbf{K}(u,v)$ indicates a mixture of temporal causality and spatial dependence between them. In inferring latent relationships in a network, we assume that entity $u$ is not related to $v$ if $\mathbf{K}(u,v)=0$. We threshold the matrix $\mathbf{K}$ at a certain level: we set elements that are smaller than the threshold value to $0$ and either maintain the values of larger or equal elements to obtain a weighted network or set them to $1$ to produce an unweighted network. We use $\mathbf{\tilde{K}}$ to denote the thresholded matrix $\textbf{{K}}$. We interpret that there is no relation between two nodes $u$ and $v$ if $\mathbf{\tilde{K}}(u,v)=\mathbf{\tilde{K}}(v,u)=0$.


\subsection{Model Estimation}\label{subsec: Estimation}

We use an EM-type algorithm \cite{veen2008estimation} to estimate the parameters and kernel functions of our model. 
This EM-type algorithm gives us an iterative method to find maximum-likelihood estimates of the parameters. 
We assume that the original model depends on unobservable latent variables. Suppose that we have data $X$ and want to estimate parameters $\Theta$. One can view the likelihood function $L(\Theta;X)$ as the marginal likelihood function of $L(\Theta;Y,X)$, where $Y$ is a latent variable.
We call $L(\Theta;Y,X)$ the ``complete-data likelihood function'' and $L(\Theta;X)$ the ``incomplete-data likelihood function''. 
Because both $Y$ and $L(\Theta;Y,X)$ are random variables, we cannot estimate them directly. Therefore, we consider the following expectation function:
\begin{align}
\label{eq:comp_like}
	Q(\Theta,\Theta^{i-1}) &= \mathbb{E}\left[\log(L(\Theta;Y,X))|X,\Theta^{i-1}\right] \nonumber\\
	&=\int \log(L(\Theta;Y,X))f(Y|X,\Theta^{i-1})\mathrm{d}Y \,,
\end{align}
where $f(Y|X,\Theta^{i-1})$ is the probability density function of $Y$, given the data $X$ and $\Theta^{i-1}$. We update parameters by solving the following equation: 
\begin{equation*}
	\hat{\Theta}^{i} = \arg\max_\Theta Q(\Theta,\Theta^{i-1})\,.
\end{equation*}


\subsubsection{Parametric Model}
\label{subsubsection: pm-em}

The log-likelihood for the parametric model defined in Equation \cref{spteH} in a spatial region $R$ and time window $[0,T]$ is
\begin{equation} \label{eq:loglike}
	\log(L(\Theta;X)) = \sum_{k=1}^{N} \log(\lambda_{u_k}(t_k)) - \sum_{u=1}^{U}\iint_R\int_0^T \lambda_u(t)\,\mathrm{d}t \,\mathrm{d}x \,\mathrm{d}y\,.
\end{equation}

We define random variables ${Y}_{ij}$ and ${Y}_{ij}^b$ using the approach from \cite{mohler2014marked}. If event $j$ triggers event $i$ via the kernel $g$, then ${Y}_{ij}=1$; otherwise, ${Y}_{ij}=0$. The equality $Y_{ij}^b=1$ indicates that event $i$ is triggered by event $j$ at a background rate of $\mu$. We define two expectation matrices $\mathbf{P}(i,j) = p_{ij} = \mathbb{E}[Y_{ij}]$ and $\mathbf{P}^b(i,j)=p_{ij}^b = \mathbb{E}[Y_{ij}^b]$. We convert the incomplete-data log-likelihood function in \cref{eq:loglike} into the following complete-data log-likelihood function:
\begin{align*}
	\log(L(\Theta;X,Y))=&\sum_{j<i} Y_{ij}\log\left(K_{u_i u_j}g(t_i-t_j,x_i-x_j,y_i-y_j)\right)-\sum_{u=1}^U\sum_{i=1}^N\beta_{uu_i}\\
	&-\sum_{u=1}^U \sum_{i=1}^N K_{u_i u}\left(1-e^{-w(T-t_i)}\right)+\sum_{i=1}^N\sum_{j=1}^N Y_{ij}^b \log(\mu_{u_i})\,.
\end{align*}
We then calculate the expectation function using \cref{eq:comp_like} to obtain
\begin{align*}
	Q(\Theta)=&\sum_{i=1}^N\sum_{j=1}^N p_{ij}^{b}\log\left(\dfrac{\beta_{u_ju_i}}{2 \pi \eta^2T} \exp{\left(-\dfrac{(x_i-x_j)^2+(y_i-y_j)^2}{2 \eta^2}\right)}\right)-\sum_{u=1}^U\sum_{i=1}^N \beta_{u_iu} \\
	&+\sum_{j<i} p_{ij}\log\left(\omega K_{ u_ju_i}  e^{-\omega (t_i-t_j)}\dfrac{1}{2\pi \sigma^2} \exp\left({-\dfrac{(x_i-x_j)^2+(y_i-y_j)^2}{2 \sigma^2}}\right)\right)\\
	&-\sum_{u=1}^U \sum_{i=1}^N K_{u_iu }\left(1-e^{-w(T-t_i)}\right)\,.
\end{align*}
We perform the maximization step of the EM-type algorithm (a projected gradient ascent) \cite{lewis2011nonparametric} directly by taking derivatives with respect to the parameters and setting them to $0$. For the expectation step, we use the ``optimal'' parameter values from the prior maximization step to update the probabilities $p_{ij}$ and $p^b_{ij}$. By (alternately) iterating these the expectation and maximization steps, we obtain \cref{alg:para} for the parametric model. For initialization, we sample $\Theta^0$, $p_{ij}$, and $p^b_{ij}$ uniformly at random. Note additionally that $p_{ij}=0$ for $i>j$.

\begin{algorithm}
	\caption{EM-type Algorithm for the Parametric Model}
	\label{alg:para}
	\begin{algorithmic}[1]
	    \STATE{\textbf{Inputs}:  
	    	point process: $\{(u_i, t_i,x_i,y_i) \}_{i=1}^{N}$; 
		initial guesses for parameters: $\Theta^{(0)}=  \left(\{K_{uv}^{(0)}\}_{u,v = 1}^U, \{\beta_{uv}^{(0)}\}_{u,v = 1}^U, \sigma^{(0)}, \omega^{(0)}\right)$ and $\{p_{ij}^{(0)}\}_{i,j=1}^N, \{p_{ij}^{b,(0)}\}_{i,j=1}^N $; termination threshold: $\epsilon$.}
		\STATE{\textbf{Outputs}: model parameters $\Theta = \left(\{K_{uv}\}_{u,v=1}^U, \{\beta_{uv}\}_{u,v = 1}^U,\sigma, \omega \right)$.}
		\STATE{Initialize $\delta = 1$ and $k=0$.}
		\WHILE{$\delta>\epsilon$}
		\STATE{Let $\eta^{2,(k)}$ and $\sigma^{2,(k)}$  be the value of $\eta^2$ and $\sigma^2$ at the $k$th iteration.} 
		\STATE{\textbf{Expectation step}: for $i,j\in\{1,2,\cdots,N\}$,}
		\STATE{$p^{(k)}_{ij} = \left(K_{ u_ju_i} g\left(t_i-t_j,x_i-x_j,y_i-y_j\right)\right)/\lambda\left(x_i,y_i,t_i\right)\,$.}
		\STATE{$p_{ij}^{b,(k)} =   \beta^{(k)}_{ u_ju_i}\exp\left(- \frac{(x_j-x_i)^2+(y_j-y_i)^2}{2\eta^{2,(k)}}\right)/2 \pi \eta^{2,(k)} T\lambda(x_i,y_i,t_i)\,$.}
		\STATE{\textbf{Maximization step}: for $u,\hat{u}\in\{1,2,\cdots,U\}$,}
		\STATE{ $\omega^{(k+1)} = \dfrac{\sum_{j<i} p_{ij}^{(k)}}{\sum_{j<i} p_{ij}^{(k)}(t_i-t_j) +\sum_{u=1}^U \sum_{i=1}^N K_{u_iu}(T-t_i)e^{-\omega(T-t_i)} }\,$,}
		\STATE{ Let $n_u$ denote the number of events in point process $u$; and let $i^u_l$, with $l \in \{1,\ldots,n_u\}$, index the events for process $u$.\\
       $K^{(k+1)}_{\hat{u}u}={\sum_{l=1}^{n_u}\sum_{t_{i^{\hat{u}}_{\hat{l}}}<t_{i^u_l}} p_{i^u_l i^{\hat{u}}_{\hat{l}}}^{(k)}}/\sum_{l=1}^{n_{\hat{u}}}\left(1-\exp{\left(-w\left(T-t_{i^{\hat{u}}_{\hat{l}}}\right)\right)}\right)\,,$}
		\STATE{
			$ \beta^{(k+1)}_{\hat{u}u}=\sum_{i=1}^{n_u}\sum_{j=1}^{n_{\hat{u}}} p_{i^u_l i^{\hat{u}}_{\hat{l}}}^{b,(k)}/n_{\hat{u}}\,$\,.  }		
		\STATE{\small
		${\sigma}^{2,(k+1)}=\sum_{i,j=1}^N \left(p_{ij}^{b,(k)}+p_{ij}^{(k)}\right)\left((x_i-x_j)^2+(y_i-y_j)^2\right)/\sum_{i,j=1}^N 2\left(p_{ij}^{b, (k)}+p_{ij}^{(k)}\right)\,$.}
	    \STATE{${\eta}^{2,(k+1)}={\sigma}^{2,(k+1)}\,$.}
		
		\STATE{$\delta=\|\Theta^{(k)}-\Theta^{(k+1)}\|$\,.}
		\STATE{$k=k+1$.}
		\ENDWHILE{}
	\end{algorithmic}
\end{algorithm}


\subsubsection{Nonparametric Model}

The log-likelihood function of the nonparametric model is the same as for the parametric model in Equation \cref{eq:loglike}. We use a similar approach as before to derive an EM-type algorithm for the nonparametric model. The main differences are that (1) only ${Y}_{ij}$ are latent variables and ${Y}_{ii}=1$ signifies that event $i$ is a background event, whereas ${Y}_{ji}=1$ signifies that event $i$ is triggered by event $j$; and (2) we assume that the triggering kernels $g_1(t)$ and $g_2(r)$ are piecewise constant functions. We discretize space and time into $n_t^{\text{bins}}$ temporal bins and $n_r^{\text{bins}}$ spatial bins, and the kernel takes a constant value in each spatiotemporal bin.

To formally present the EM-type algorithm (see \cref{alg:nonpara}), we borrow notation from \cite{fox2016spatially}. Let $C_k$ denote the set of event pairs $(i,j)$ for which $t_j-t_i$ belongs to the $k^{\mathrm{th}}$ temporal bin, $D_k$ denote the set of event pairs $(i,j)$ for which $r_{ij}$ (the distance between nodes $i$ and $j$) belongs to the $k^{\mathrm{th}}$ spatial bin, $N_{u}$ denote the number of events that include node $u$, the parameter $\Delta t_k$ denote the size of the $k^{\mathrm{th}}$ temporal bin, and $\Delta r_k$ denote the size of the $k^{\mathrm{th}}$ spatial bin.

\begin{algorithm}
	\caption{EM-type Algorithm for our Nonparametric Model}
	\label{alg:nonpara}
	\begin{algorithmic}[1]
	\STATE{\textbf{Inputs}: point process: $\{(u_i, t_i,x_i,y_i)\}_{i=1}^{N}$; initial guesses of parameters: $ \{K_{uv}^{(0)}\}_{u,v = 1}^U$ and $\{p_{ij}^{(0)}\}_{i,j=1}^N$; termination threshold: $\epsilon$.}
	\STATE{\textbf{Outputs}: model parameters: $\{K_{uv}\}_{u,v=1}^U$; triggering probability between events: $\{p_{ij}\}_{i,j=1}^N$; temporal triggering kernel: $g_1$; spatial triggering kernel: $g_2$.}
		\STATE{Initialize $\delta = 1$ and $\eta=0$\,.}
		\WHILE{$\delta>\epsilon$}
		\STATE{Update background kernel $\tau^{\eta}(x,y)$ (see \cref{eq:nonpara_cond})}
		\STATE{ $\gamma_u^{(\eta)}=\sum_{u_i=u} p^{(\eta)}_{ii}/Z^{(\eta)}$\,, where $Z^{(\eta)}$ satisfies $\int_0^T \iint_S \tau^{\eta}(x,y)\mathrm{d}s\,\mathrm{d}t=Z^{(\eta)}$ for a bounded spatial domain $S$ and for $u \in \{1,\ldots,U\}$.}
		\STATE{ $K^{(\eta)}_{uv}=\sum_{u_i=u}\sum_{u_j=v}p^{(\eta)}_{ij}/N_{u}$ for $u,v \in \{1,\ldots,U\}$.}
		\STATE{$g_1^{(\eta)}(t)=\sum_{i,j \in C_k}p^{(\eta)}_{ij}/\Delta t_k \sum_{i<j}p^{(\eta)}_{ij}$ for $t$  in the $k^\mathrm{th}$ temporal bin.}
		\STATE{$h^{(\eta)}(r)=\sum_{i,j \in D_k} p^{(\eta)}_{ij}/\Delta r_k \sum_{i<j}p^{(\eta)}_{ij}$ for $r$ in the $k^\mathrm{th}$ spatial bin. Set $g_2^{(\eta)}(r) =  h^{(\eta)}(r)/(2\pi r)$\,.}
		\STATE{ $p^{(\eta+1)}_{ij}=K^{(\eta)}_{u_iu_j}g_1^{(\eta)}(t_j-t_i)g_2^{(\eta)}(r_{ij})$ for $i<j$ and $p^{(\eta+1)}_{jj}=\mu^{(\eta)}_{u_j}(x_j,y_j)$.} 
		\STATE {Normalize $p_{ij}^{(\eta+1)}$ so that $\sum_{i=1}^N p_{ij}^{(\eta+1)}=1$ for any $j$.}
		
		\STATE{$\delta=\max_{i,j} \|p^{(\eta+1)}_{ij}-p^{(\eta)}_{ij} \|$ and $\eta=\eta+1$.}
		\ENDWHILE
	\end{algorithmic}
\end{algorithm}


\subsection{Simulations}

To generate synthetic data for model comparisons, we need to simulate self-exciting point processes with the conditional intensity in Equation \cref{spteH} for each process $u$. We use the branching structures \cite{zhuang2004analyzing} of self-exciting point processes to develop \cref{alg:simu} for our simulations.

\begin{algorithm}
	\caption{Simulation of a Multivariate Hawkes Process}
	\label{alg:simu}
	\begin{algorithmic}[1]
	    \STATE{\textbf{Inputs}: time-window size: $T$; spatial region: $S\subset \mathbb{R}^2$; background rate: $\{\gamma_u\}_{u=1}^U$; triggering matrix: $\{K_{uv}\}_{u,v=1}^U$; temporal and spatial triggering kernels: $g_1(t)$, $g_2(x,y)$\,.}
	    \STATE{\textbf{Output}: point process: $\mathbf{C}=\{(u_i, t_i,x_i,y_i) \}_{i=1}^{N}$\,.}
	    \STATE{Initialize an empty set $\mathbf{C}$ and an empty stack $\mathbf{Q}$.}
		\STATE{\textbf{Generate background events:}}
		\begin{ALC@g}
		    \STATE{Draw $N_u$, the number of background events of type $u$, from a Poisson distribution with parameter $\lambda=\gamma_uT$ for each $u\le U$.}
		    \STATE{Add each background event $i\le \sum_{u=1}^U N_u$ --- i.e., $(x_i,y_i,t_i,u_i)$ --- to the set $\mathbf{C}$ and the stack $\mathbf{Q}$, where $(x_i,y_i,t_i)$ is drawn from the uniform spatiotemporal distribution over the time interval $[0,T]$ and a bounded spatial region $S$.}
		\end{ALC@g}
		\STATE{\textbf{Generate triggered events:}}
		\begin{ALC@g}
		    \WHILE{$\mathbf{Q}$ is not empty}
		        \STATE{Remove the most recently added element $(x_i,y_i,t_i,u_i)$ from the stack $\mathbf{Q}$.} 
		        \STATE{Draw $N_i$, the number of events triggered by event $i$, from a Poisson distribution with parameter $\lambda_i=\sum_{u'=1}^U K_{u_iu'}$.}
		        \STATE{Generate events $(x_k,y_k,t_k,u_k)$ for each $k\le N_i$ as follows:}
		        \begin{ALC@g}
		            \STATE{Sample $t_k$, $(x_k,y_k)$ and $u_k$ according to $g_1(t-t_i)$, $g_2(x-x_i,y-y_i)$, and $P(u_k=\tilde{u})=\frac{K_{u_i\tilde{u}}}{\sum_{v=1}^U K_{u_iv}}$, respectively.}
		            \STATE{Add $(x_k,y_k,t_k,u_k)$ to the set $\mathbf{C}$.}
		            \IF{$t_k\le T$}
		                \STATE{Add the element $(x_k,y_k,t_k,u_k)$ to the stack $\mathbf{Q}$.}
		            \ENDIF
		        \end{ALC@g}
		    \ENDWHILE
		\end{ALC@g}
	\end{algorithmic}
\end{algorithm}


\section{Numerical Experiments and Results}\label{sec:experiments}

We apply our algorithm to both synthetic and real-world data sets to demonstrate the usefulness of incorporating spatial information and of our nonparametric approach. We consider a synthetic data set in \cref{subsec:syn}, a Gowalla data set in \cref{sec:Gowalla}, a crime-topic network data set in \cref{sec:CTN}, and a violent gang-crime data set in \cref{sec:NCE}. Using the first three of these data sets, we compare our nonparametric model (``Nonparametric Hawkes'') with the Bayesian Hawkes model\footnote{We use code from the authors of \cite{linderman_discovering_2014}; it is available at \protect~\url{https://github.com/slinderman/pyhawkes}. In all of our experiments, we use the default hyperparameters that come with the published code.} in \cite{linderman_discovering_2014} (``Bayesian Hawkes''), the exclusively temporal Hawkes model with kernel $g(t) = \omega \exp(-\omega t)$ from \cite{fox_modeling_2016} (``Temporal Hawkes''), and the parametric spatiotemporal model detailed in \cref{subsec: a parametric model} (``Parametric Hawkes''). We make comparisons by examining how well the following properties are recovered in the inferred triggering matrix: (1) symmetry and reciprocity; (2) existence of edges; and (3) community structure. We also demonstrate the ability of our algorithm to infer the triggering kernel $g$. Using the fourth data set (see \cref{sec:NCE}), we study a network of crime events using a violent gang-crime data set. We examine relations between crime events and repeated triggering patterns.


\subsection{Synthetic Data} \label{subsec:syn}


We first generate synthetic triggering matrices $\mathbf{K}$ using a weighted stochastic block model (WSBM) \cite{aicher_learning_2015, peixoto2017bayesian}. We assign a network's nodes to four sets (called ``communities'') and assign edges to adjacency-matrix blocks based on the set memberships of the nodes. Two of the communities consist of ten nodes each, and the other two communities consist of five nodes each. For each edge, we first draw a Bernoulli random variable to determine whether it exists, and we then draw an exponential random variable to determine the weight of the edge (if it exists). The parameter of the Bernoulli random variable is $0.68$ for there to be an edge between nodes from the same community and $0.2$ for an edge between nodes from different communities. The decay-rate parameter for the exponential random variable in these two situations is $0.1$ and $0.01$, respectively. By construction, our triggering matrices are symmetric. 

The triggering matrices that we generate in this way are not guaranteed to satisfy the stability condition for Hawkes processes; this condition is that the largest-magnitude eigenvalue of $\mathbf{K}$ is smaller than one \cite{daley2007introduction}. When this condition is satisfied, each event has, almost surely, finitely many subsequent events as ``offspring''. In our work, we discard any simulated adjacency matrix that does not satisfy the stability condition, and we generate a new one to replace it. (With our choices of the parameters, we discard about 65\% of the generated adjacency matrices.)

With each triggering matrix $\mathbf{K}$, we use \cref{alg:simu} to simulate a multivariate spatiotemporal Hawkes process with our parametric model in \cref{subsec: a parametric model} with $\omega = 0.6$, $\sigma^2=0.3$, $T=250$, $S = [0,1]\times[0,1]$, and a homogeneous value $\gamma_u = 0.2$ for all nodes $u$. We then reconstruct the underlying networks and the triggering kernels from the simulated data. 


\subsubsection{Symmetry and Reciprocity}

As we noted in \cref{subsec:syn}, our simulated triggering matrices are symmetric, but our reconstructed adjacency matrices generally are not symmetric. Measuring deviation from symmetry gives one way to evaluate the performance of our inference methods. We use various reciprocity measures to quantify such deviation. 

We conduct two sets of experiments. In the first one, we fix a single synthetic triggering matrix and simulate ten multivariate spatiotemporal Hawkes point processes. We then estimate the triggering matrix $\mathbf{K}$ from each point process using various methods, which we thereby compare with each other. In a second set of experiments, instead of fixing a single triggering matrix, we generate ten different triggering matrices using the same WSBM model and parameters, and we simulate one point process for each triggering matrix. 

There is no standard way of measuring reciprocity in a weighted network. In our calculations, we use diagnostics that were proposed in \cite{squartini2013reciprocity} and \cite{akoglu2012quantifying}. First, as in \cite{squartini2013reciprocity}, we compute the reciprocated edge weight $K_{uv}^{\leftrightarrow} = \min\{K_{uv}, K_{vu}\}$, and we then calculate a network-level reciprocity score $R_1$ as the ratio between the total reciprocated weight $W^{\leftrightarrow} = \sum_{u\neq v}K_{uv}^{\leftrightarrow}$ and the total weight $W = \sum_{u\neq v}K_{uv}$. That is, the ``reciprocity'' is $R_1 := {W^\leftrightarrow}/{W}$. Second, Akoglu et al.~\cite{akoglu2012quantifying} proposed three node-level measures of reciprocity: (1) the ``ratio'' $R_\mathrm{ratio} := \min\{K_{uv},K_{vu}\}/\max\{K_{uv},K_{vu}\}$; (2) ``coherence'' $R_\mathrm{coher} = 2\sqrt{K_{uv}K_{vu}}/(K_{uv} + K_{vu})$; and (3) ``entropy'' $R_\mathrm{entropy} := -r_{uv}\log_2(r_{uv}) - r_{vu}\log_2(r_{vu})$, where $r_{uv} = K_{uv}/(K_{uv}+K_{vu})$. These last three measures of reciprocity are measured at a node level, whereas $R_1$ is a network-level measure. For the other measures, we obtain a network-level measure by calculating those scores for each pair of nodes and then taking a mean over all pairs of nodes. Each of the above quantities gives a score between $0$ and $1$, where a larger value indicates a stronger tendency for the nodes in a network to reciprocate. In a perfectly symmetric and reciprocal network, each of the four methods gives a value of $1$.

In \cref{table:rec}, we report the mean reciprocity and the standard deviation over ten simulations with the same triggering matrix. In \cref{table:rec2}, we report the mean results from ten different triggering matrices. Both spatiotemporal models give higher scores than the exclusively temporal models, which is what we expected, as the temporal models discard spatial information. According to these measures of success, the nonparametric model has the best performance.

\begin{table}
	\centering
	\footnotesize{
		\caption{Reciprocity of the triggering matrices that we infer using different methods. We report the mean and standard deviation (in parentheses) over ten simulations with the same (ground-truth) triggering matrix.\label{table:rec}}
		\begin{tabular}{|c|c|c|c|c|}
			\hline
			& Nonparametric & Temporal & Parametric  & Bayesian\\\hline
			$R_1$ & {0.59} (0.05) & 0.29 (0.06) & 0.54 (0.03)& 0.36 (0.03) \\\hline
			Correlation & {0.84} (0.05) & 0.36 (0.16) & 0.79 (0.05) & 0.30 (0.14) \\\hline
			Ratio & 0.55 (0.02) & 0.37 (0.11) & {0.58} (0.02) & 0.32 (0.02)\\\hline
			Coherence & {0.75} (0.01) & 0.63 (0.03) & 0.71 (0.02) &0.68 (0.02) \\\hline
			Entropy & {0.71} (0.01) & 0.59 (0.03) & 0.68 (0.02) & 0.60 (0.02) \\\hline
		\end{tabular}
	}
\end{table}
\begin{table}
	\centering
	\footnotesize{
		\caption{Reciprocity of the triggering matrices that we infer using different methods. We report the mean and standard deviation (in parentheses) over ten simulations, each with a different (ground-truth) triggering matrix.
		\label{table:rec2}}
		\begin{tabular}{|c|c|c|c|c|}
			\hline
			& Nonparametric & Temporal & Parametric  & Bayesian \\\hline
			$R_1$ & {0.61} (0.12) & 0.36 (0.12) & 0.55  (0.10) & 0.40 (0.05) \\\hline
			Correlation & {0.81} (0.16) & 0.48 (0.27) & 0.76 (0.15) & 0.23 (0.14) \\\hline
			Ratio & {0.63} (0.04) & 0.43 (0.06) & 0.62 (0.03) & 0.33 (0.03)\\\hline
			Coherence & {0.78} (0.04) & 0.62 (0.03) & 0.72 (0.03)& 0.70 (0.03) \\\hline
			Entropy & {0.75} (0.05) & 0.58 (0.03) & 0.69 (0.03) & 0.62 (0.04) \\\hline
		\end{tabular}
	}
\end{table}


\subsubsection{Edge Reconstruction} \label{sec:spr}

We also evaluate the reconstruction methods based on their ability to recover the existence of edges. This is particularly relevant if we want to know whether there is a connection between two entities. We will discuss this application in detail using the Gowalla data set (see \cref{sec:Gowalla}). 

\begin{figure}	
	\centering
	\includegraphics[width=0.5\linewidth]{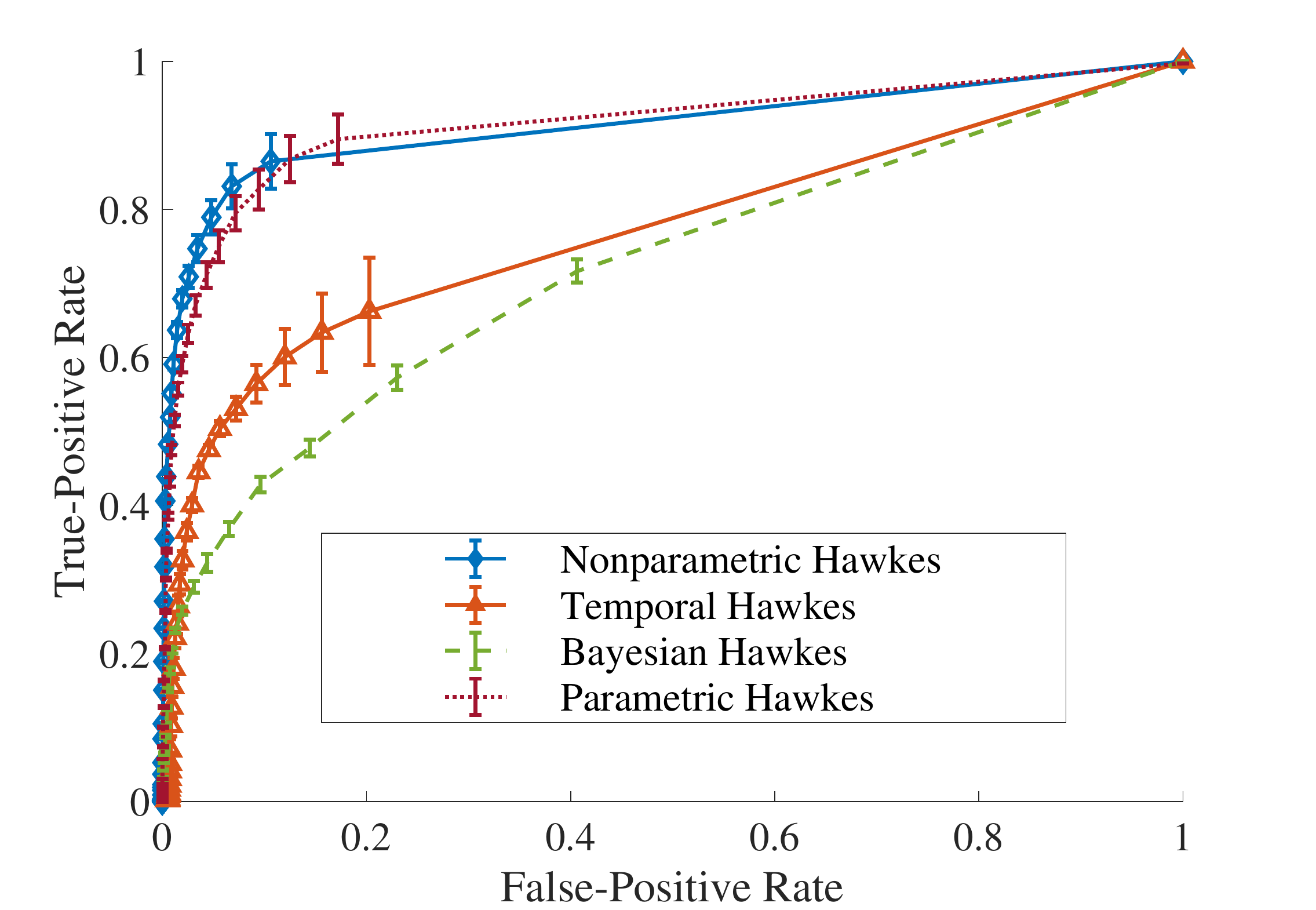}
	\caption{Model comparison using synthetic networks. We show the mean ROC curves with error bars (averaged over ten simulations, each with a different triggering matrix) on edge reconstruction. The ROC curve of a better reconstruction should be closer to $1$ for a larger range of horizontal-axis values, such that it has a larger area under the curve (AUC), which is equal to the probability that a uniformly-randomly chosen existing edge in a ground-truth network has a larger weight than a uniformly-randomly chosen missing edge in the inferred network.
	}	
	\label{fig:test}
\end{figure}

In our model, we consider an edge to exist if the corresponding weighted entry in the inferred triggering matrix exceeds a certain threshold. For different threshold levels, we compute the numbers of true positives (TP), false positives (FP), true negatives (TN), and false negatives (FN) for a given ground-truth triggering matrix. We summarize our results in a receiver operating characteristic (ROC) plot (see \cref{fig:test}), in which we plot the true-positive rate (TPR) (where $\text{TPR} = \text{TP}/(\text{TP} + \text{FN})$) versus the false-positive rate (FPR) (where $\text{FPR} = \text{FP}/(\text{FP} + \text{TN})$). A better inference of a triggering matrix gives a larger value of TPR at a fixed FPR. 

Based on the ROC plot in \cref{fig:test}, we conclude that the spatiotemporal models --- both the parametric and nonparametric Hawkes models that we proposed in \cref{sec:alg} --- outperform the exclusively temporal ones. Therefore, incorporating spatial information improves the quality of our reconstructed binary networks, at least according to this measure of success. The best results are from our parametric model, which is not surprising, given that we use the same model to simulate the data. The performance of our nonparametric model is very close to that of the parametric model, confirming its effectiveness at inferring the existence of edges.

\subsubsection{Inferred Kernels}

\begin{figure}
	\centering
	\includegraphics[width=.8\linewidth]{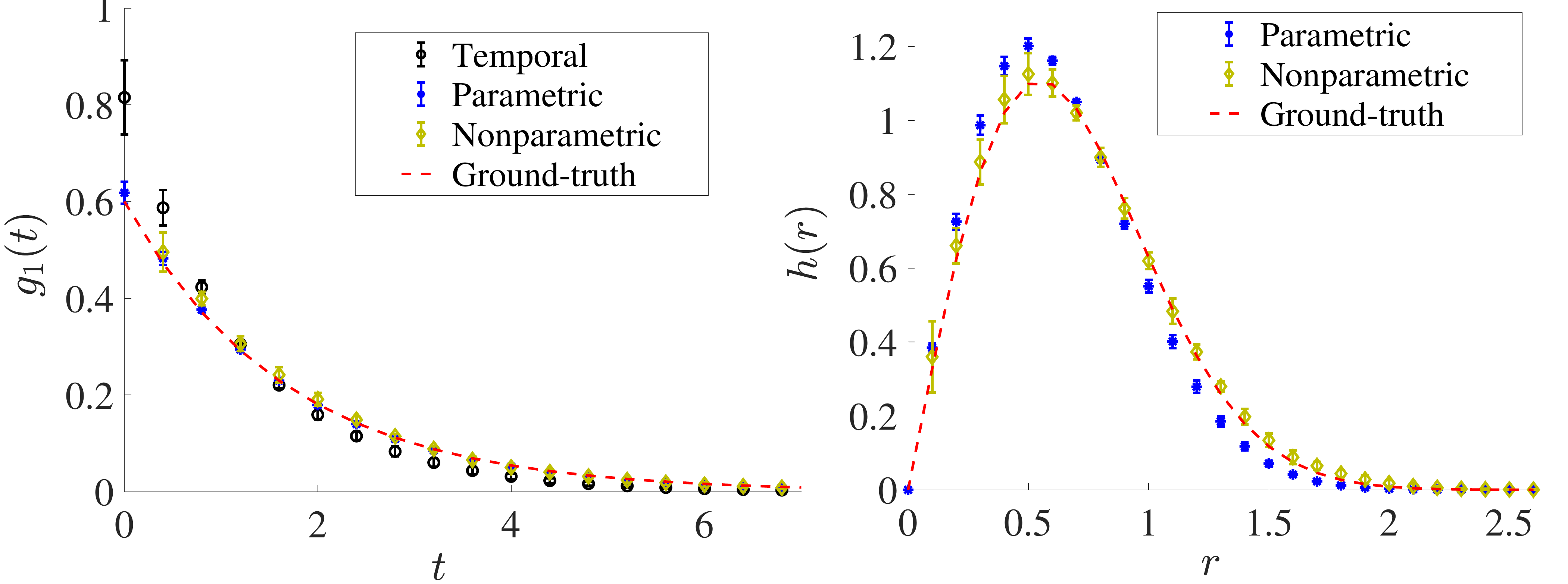}
	\caption{Model comparison using synthetic networks: 
	Inferred (left) temporal and (right) spatial kernels using different methods: Temporal Hawkes, Parametric Hawkes and Nonparametric Hawkes. The dashed lines are ground-truth kernels used for the synthetic data.
	}
	\label{fig:kernel}
		
\end{figure}

We report the inferred kernels of the different models in \cref{fig:kernel}. Recall that the ground-truth kernels that we use to simulate point processes are $g_1(t) =\omega \exp\left(-\omega t\right) $ and $h(r) = 2\pi rg_2(r)= \frac{r}{ \sigma^2} \exp \left(-\frac{r^2}{2 \sigma^2}\right)$, where $r^2 = x^2 + y^2$, $\omega = 0.6$, and $\sigma^2 = 0.3$. Let $\hat{g_1}$ and $\hat{h}$ denote the inferred temporal and spatial kernels, respectively. 

We calculate the $L_1$ errors $\int |g_1(t) - \hat{g_1}(t)| \,\mathrm{d}t$ and $\int |h(r)-\hat{h}(r)|\,\mathrm{d}r\,$.
We report these errors in \cref{tab:kernel} and present visualizations of the inferred kernels in \cref{fig:kernel}. 
As expected, both spatiotemporal Hawkes models give more accurate kernel inference than the exclusively temporal model. The nonparametric Hawkes model does not use any information about the ground-truth kernels. Surprisingly, it is more accurate, in terms of the $L_1$ error, at inferring the spatial trigger kernel than the parametric model, whose kernel shares the same parametric form as the ground-truth kernel.

\begin{table}
\centering
{\footnotesize 
    \caption{The $L_1$ errors of the inferred spatial and temporal kernels. We simulate ten point processes with the same triggering matrix and triggering kernel. We report the mean and standard deviation (in parentheses) of the $L_1$ errors averaged over the ten simulations with the same triggering kernel and matrix. Note that the exclusively temporal model does not estimate a spatial kernel.}
    \begin{tabular}{|c|c|c|c|}
    \hline
         & Nonparametric & Temporal & Parametric  \\\hline
         Temporal kernel & 0.07 (0.02) &0.20 (0.06) & 0.02 (0.02) \\\hline
         Spatial kernel &0.06 (0.02)& -&0.12 (0.02)\\\hline
    \end{tabular}
        \label{tab:kernel}
 }
\end{table}

\subsubsection{Community-Structure Recovery} \label{sec:community}

We also evaluate the quality of the inferred networks based on their community structure, in which dense sets of nodes in a network are connected sparsely to other dense sets of nodes \cite{porter_communities_2009,fortunato_community_2016}. Recall that we have planted a four-community structure in the synthetic triggering matrices (see \cref{subsec:syn}).
We apply the community-detection methods from \cite{aicher_learning_2015} (an inference method for a WSBM), \cite{kuang_symmetric_2012} (symmetric non-negative matrix factorization; NMF), and  \cite{jutla_generalized_2011,newman_finding_2004,newman_finding_2006,mucha2010community} (modularity maximization\footnote{For modularity maximization, we use the implementation of a (locally greedy) Louvain-like \cite{blondel2008} method (called {\sc GenLouvain}) from \cite{jutla_generalized_2011} with the default resolution-parameter value of $1$ and the Newman--Girvan null model.}). The WSBM that we infer for community detection is the same that one we use to construct the synthetic adjacency matrices (see \cref{subsec:syn}). To evaluate our inferred community structure, we use the square-root variant of \emph{normalized mutual information} (NMI) \cite{strehl2002cluster} between the inferred community assignment and ``ground truth'' community labels. Specifically, Let $S_1$ and $S_2$ be community assignments of the $U$ nodes to $C_1$ and $C_2$ communities, respectively; and let $S_{\ell k}$, with $\ell\in \{1,2\}$ and $k\in \{1,2,\cdots,C_\ell\}$, denote the set of nodes in the $k$th community in assignment $S_\ell$. The NMI between $S_1$ and $S_2$ is
\begin{equation*}
	\mathrm{NMI}(S_1,S_2) = \frac{I(S_1,S_2)}{\sqrt{H(S_1)H(S_2)}} \in [0,1]\,,
\end{equation*}
where $I(S_1,S_2) = \sum_{i=1}^{C_1}\sum_{j=1}^{C_2} \frac{|S_{1i}\cap S_{2j}|}{U}\log\frac{|S_{1i}\cap S_{2j}|/U}{|S_{1i}||S_{2j}|/U^2}$ (where $|J|$ denotes the cardinality of the set $J$) and the entropy is $H(S_\ell) = - \sum_{i=1}^{N_{\ell}} \frac{|S_{\ell i}|}{N}\log \frac{|S_{\ell i}|}{N}$ (with $\ell\in \{1,2\}$). Intuitively, NMI measures the amount of information that is shared by two community assignments. If they are the same after permuting community labels, the NMI is equal to $1$. A larger NMI score implies that the inferred community assignment shares more information with the ground-truth labels. See \cite{traud2011} for a discussion of other approaches for comparing different community assignments in networks.

There are numerous approaches for detecting communities in networks  \cite{fortunato_community_2016,porter_communities_2009,peixoto2017bayesian}, and we use methods with readily-available code. As we show in \cref{table:sbm}, all of these community-detection methods perform better when we infer triggering matrices using both spatial and temporal information than with with exclusively temporal information. One can, of course, repeat our experiments using other methods.

\begin{table}
	\centering
	\footnotesize{
		\caption{Normalized mutual information (NMI) between the outputs of different community-detection methods applied to the 
		inferred networks and the ground-truth community structure (averaged over ten simulations, each with a different triggering matrix).
		\label{table:sbm}}
		\begin{tabular}{|c|c|c|c|c|}
			\hline
			& Nonparametric & Temporal & Parametric   & Bayesian\\\hline
			Weighted SBM & 0.80 & 0.38 & {0.83} & 0.36 \\\hline
			Symmetric NMF & 0.62& 0.31 &{0.66}& 0.19\\\hline
			Modularity Maximization& 0.64 & 0.47 & {0.71} & 0.28\\\hline
		\end{tabular}
	}
\end{table}


\subsection{Gowalla Friendship Network} \label{sec:Gowalla}

Gowalla is a location-based social-media website in which users share their locations by checking in. We use a Gowalla data set --- collected in \cite{cho_friendship_2011} using Gowalla's public API --- of a ``friendship'' network with 196,591 users, 950,327 edges, and a total of 6,442,890 check-ins of these users between February 2009 and October 2010. The data set also includes the latitude and longitude coordinates and the time (with a precision of one second) of each check-in. Similar to a Facebook ``friendship'' network, the Gowalla friendship network is undirected.  
The mean number of friends for each user is $9.7$, the median is $3$, and the maximum is $14,730$. We study several subnetworks in the Gowalla data set; see \cref{appA} for details. We view the spatiotemporal check-ins of Gowalla users within each subnetwork as events in a multivariate point process and infer relationships between these users. 

\begin{figure}
	\centering
	\subfloat[San Fransisco]{\includegraphics[width=0.333\textwidth]{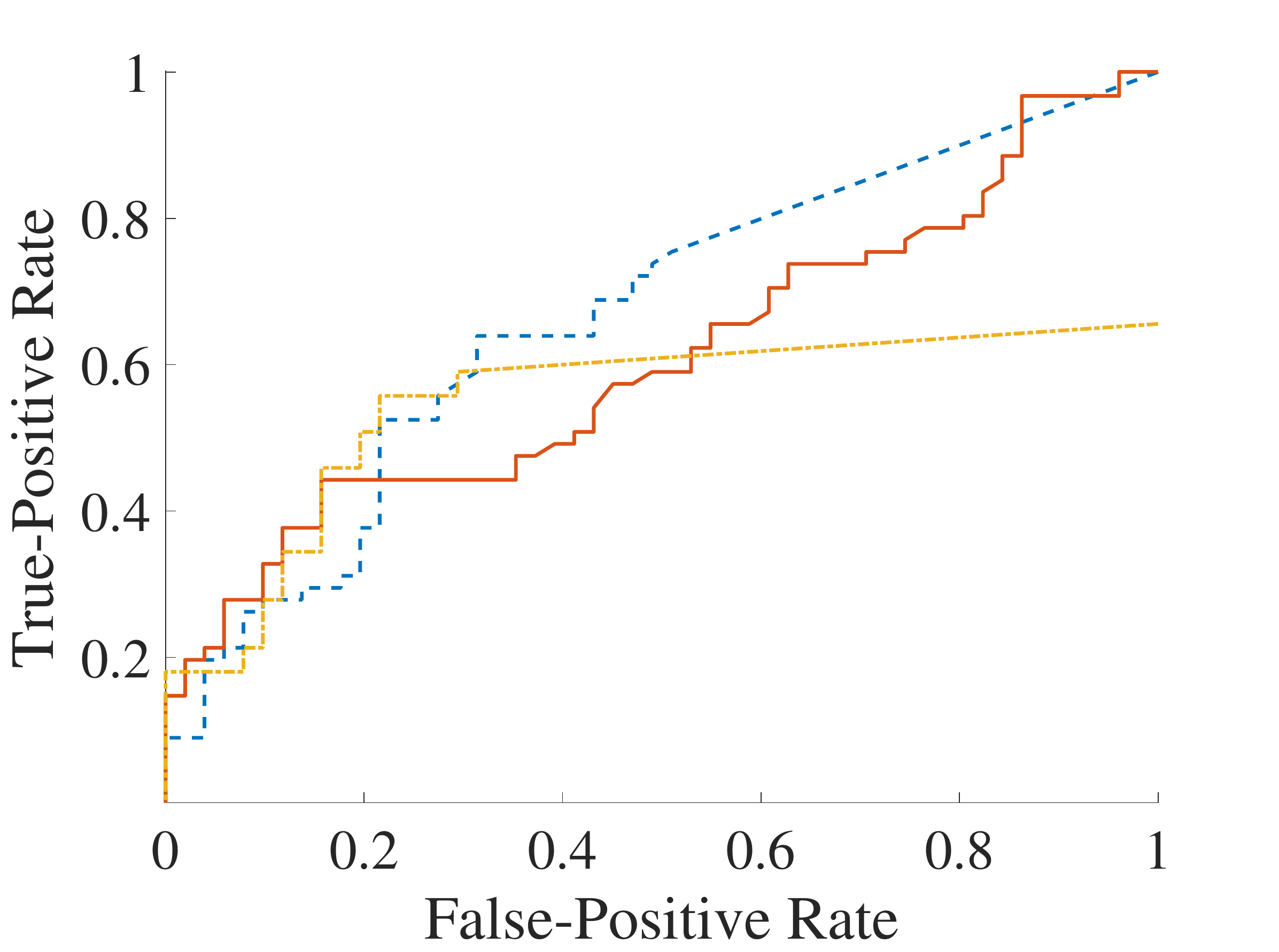}}
	\subfloat[New York City]{\includegraphics[width=0.333\textwidth]{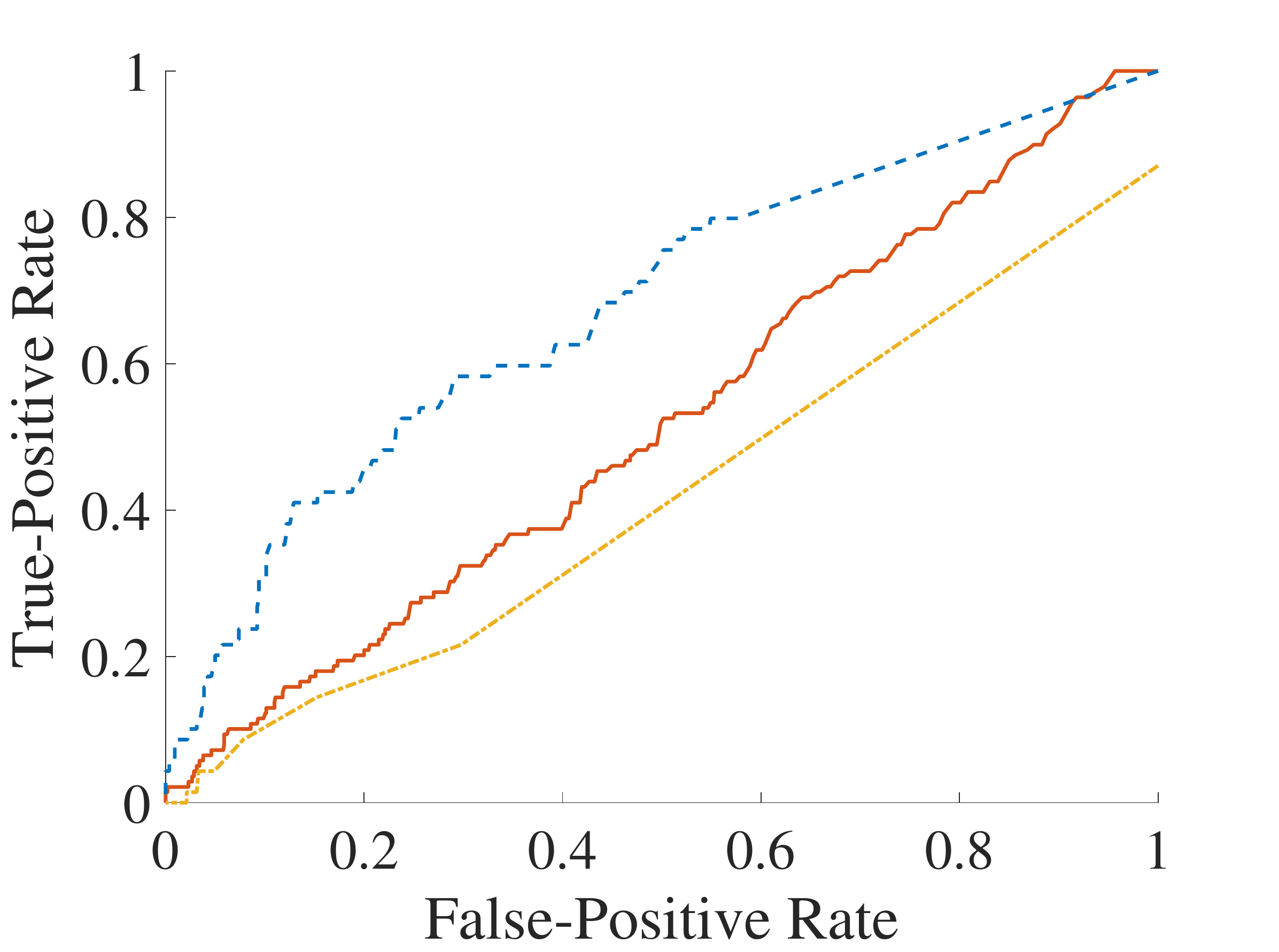}}
	\subfloat[Los Angeles]{\includegraphics[width=0.333\textwidth]{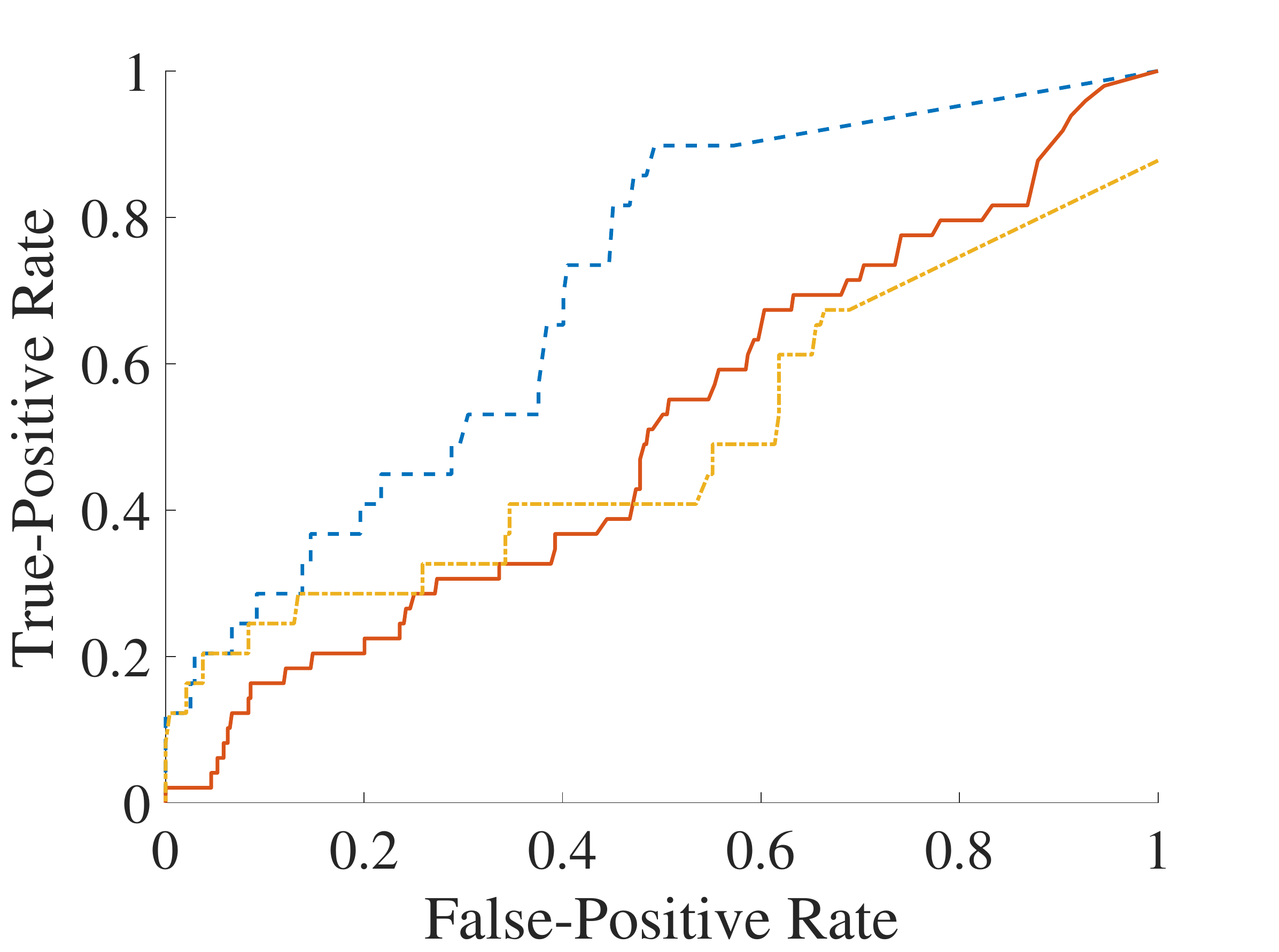}}
	\caption{ROC curves of different methods for reconstructing three Gowalla friendship networks. (See \cref{appA} for details about the networks.) Here dashed lines are for our Nonparametric Hawkes; dotted lines for Temporal Hawkes; and solid lines for Bayesian Hawkes.)
	}
	\label{fig:gow_result}
\end{figure}

We compare our Nonparametric Hawkes method with the Bayesian Hawkes and the exclusively Temporal Hawkes in terms of how well our inferred edges match the Gowalla friendships. Because a Gowalla friendship network is undirected, we first symmetrize the inferred triggering matrix (via $\tilde{\mathbf{K}} = \left(\mathbf{K} + \mathbf{K}^T\right)/2$) to obtain an undirected network. We then calculate FPRs and TPRs in the same fashion as \cref{sec:spr} using $\tilde{\mathbf{K}}$'s associated ``ground-truth'' friendship network and generate the corresponding ROC curves. In the ROC curves of three different cities in \cref{fig:gow_result}, we observe that the best results are from our nonparametric model that incorporates spatial information. The mean AUCs are $0.4277$ (with a standard deviation of $0.1042$) for the Temporal Hawkes method; $0.5301$ (with a standard deviation of $0.0585$) for the Bayesian Hawkes method; and $0.6692$ (with a standard deviation of $0.0421$) for our Nonparametric Hawkes method in all of the examined subnetworks.

\subsection{Crime-Topic Network} \label{sec:CTN}

In a recent paper on crime classification, Kuang et al.~\cite{kuang_crime_2017} performed topic modeling (see \cite{kuang2015nonnegative} for a review) on short narrative (i.e., text) descriptions of all crimes, with spatial coordinates and time stamps (with a precision of a minute), that were reported to the Los Angeles Police Department (LAPD) between 1 January 2009 and 19 July 2014. The premise in their work was that crime topics, sets of words that co-occur frequently in the same crime narrative, better reflect the ecological circumstances of crime than standard crime classifications based on legal codes. Targeting discovery of up to twenty topics, they found six topics related to violent crime, eight topics related to property crime, and six topics that seem to be related to deception-based crime. This classifies the twenty crime topics into three classes. 

In the present case study, we extend this work by modeling the above data set as a crime-topic network. We associate each crime topic with a node, and we infer edges based on whether crime events of one topic trigger events of other topics. That is, we discover latent relationships between different crime topics based on associated crime events. Inspired by previous research on point-process models of crime events \cite{mohler_self-exciting_2011}, we model crime events of different topics via a multivariate point process and infer connections between the crime topics using our Nonparametric Hawkes method. To evaluate our approach, we compare the communities that we detect in the reconstructed network with the three crime classes in \cite{kuang_crime_2017}. 
 

\subsubsection{Community Detection}

\begin{figure}
		\centering 
		(a)\subfloat{\includegraphics[width=0.18\linewidth]{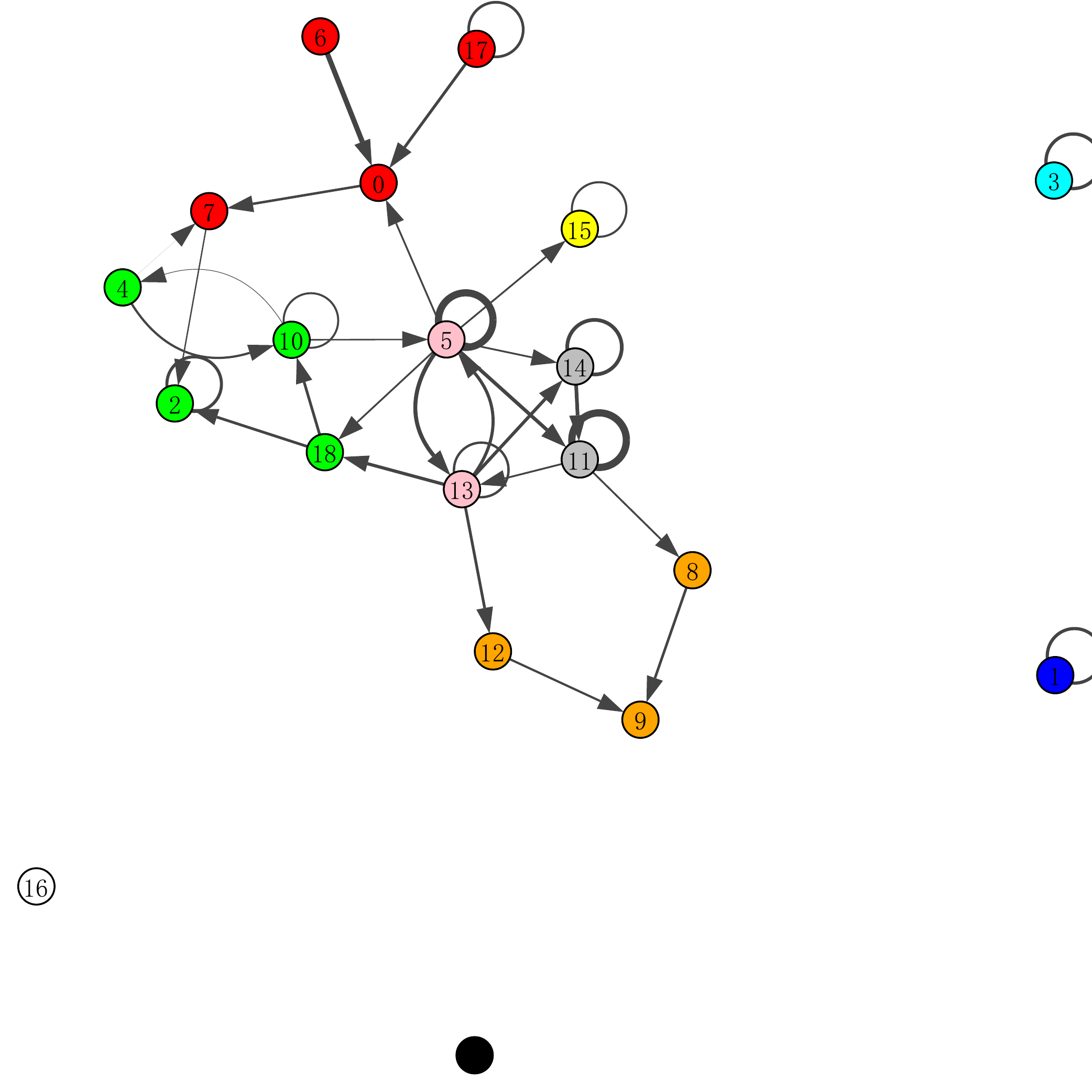} } \hspace{2mm}
		(b)\subfloat{\includegraphics[width=0.18\linewidth]{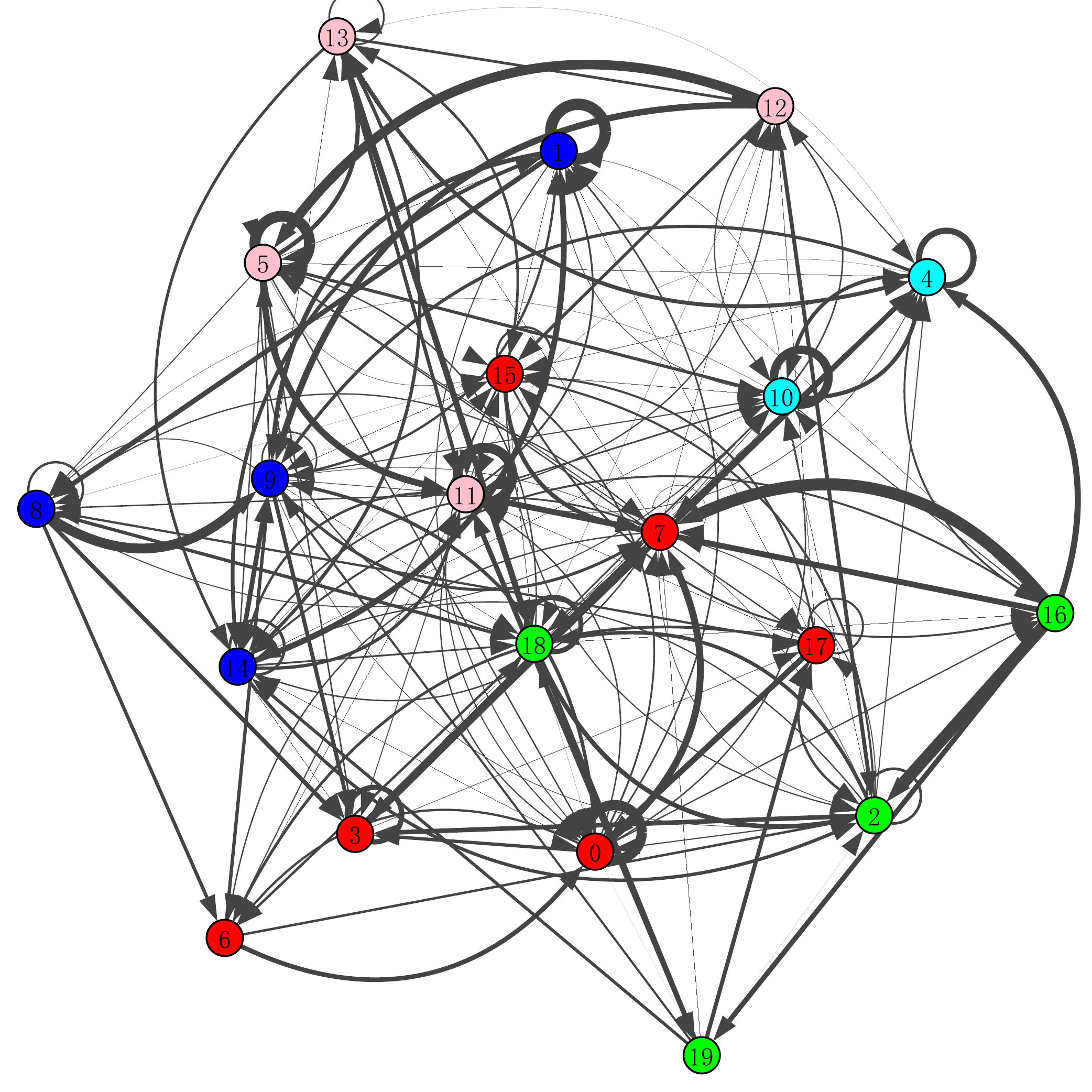} } \hspace{2mm}
		(c)\subfloat{\includegraphics[width=0.18\linewidth]{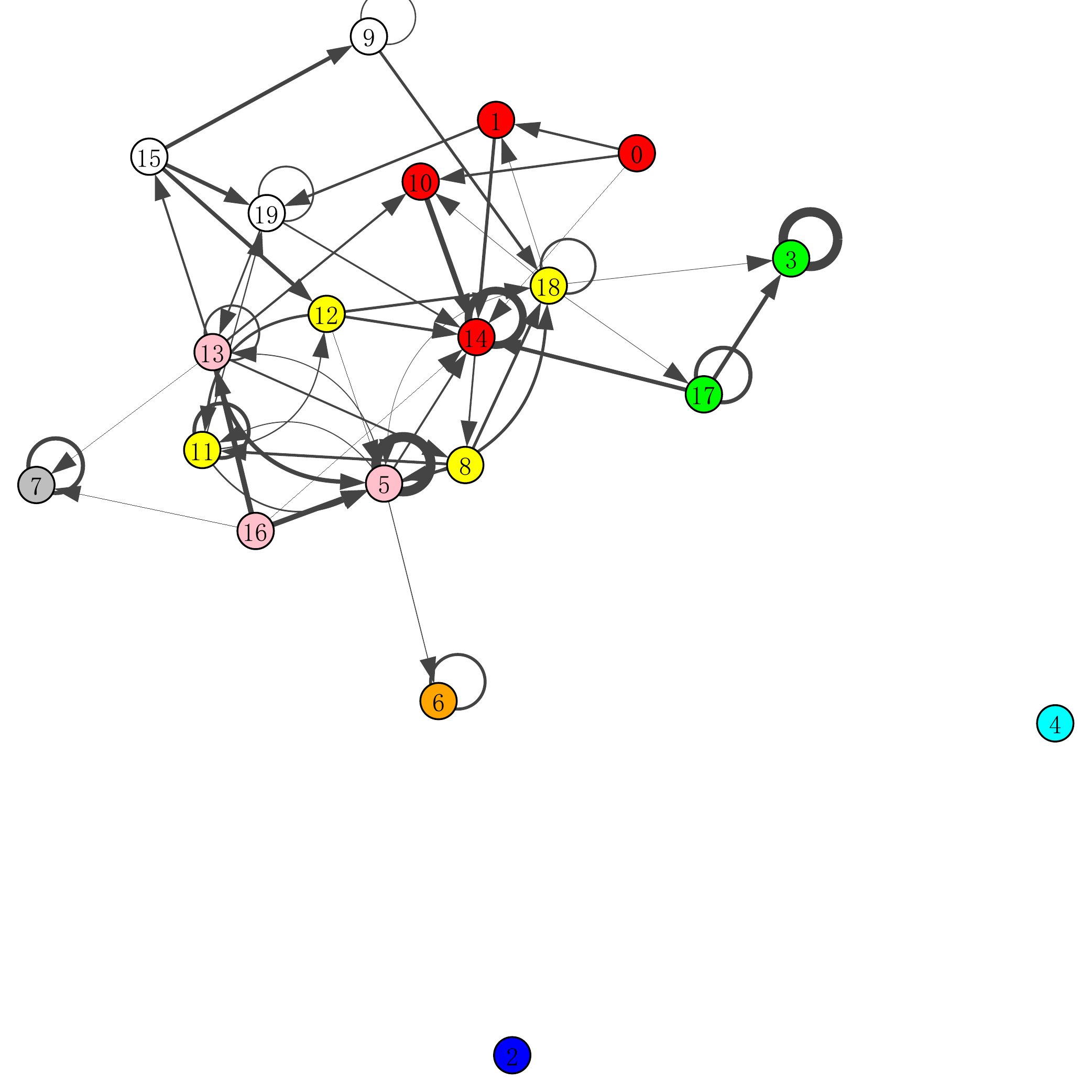} } \hspace{2mm}
		(d)\subfloat{\includegraphics[width=0.18\linewidth]{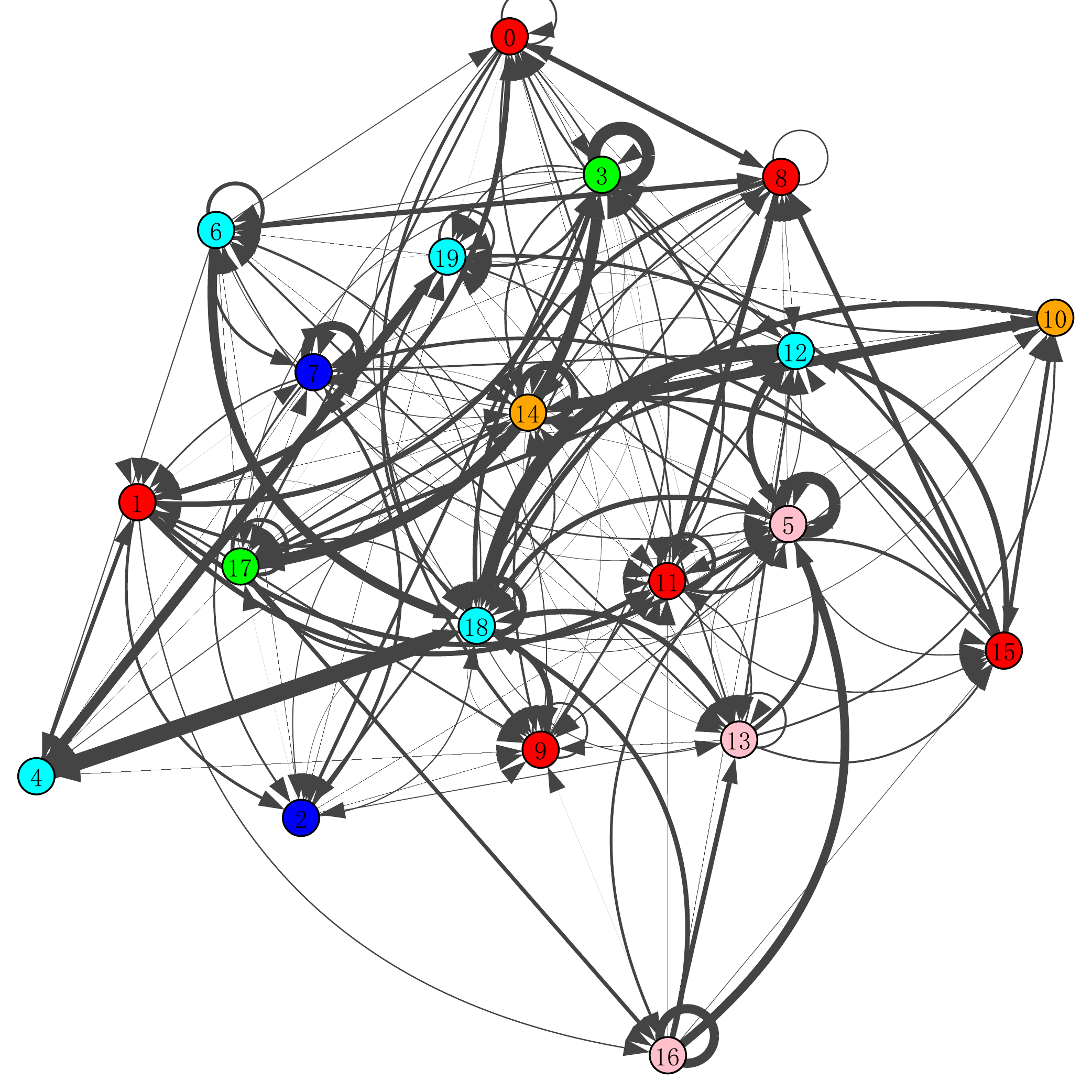}}
	\caption{Crime-Topic networks generated by the Nonparametric Hawkes and Temporal Hawkes methods colored by community assignments from modularity maximization: (a) Nonparametric Hawkes in Westwood, (b) Temporal Hawkes in Westwood, (c) Nonparametric Hawkes in Wingfoot, and (d) Temporal Hawkes in Wingfoot.
	}
	\label{fig:westwood}
\end{figure}

We infer crime-topic networks directly from crime events within individual Los Angeles neighborhoods\footnote{We use the Zillow neighborhood boundaries from \protect~\url{https://www.zillow.com/howto/api/neighborhood-boundaries.htm}.} using our Nonparametric Hawkes method, the Parametric Hawkes method, and the exclusively temporal Hawkes method. We investigate the 100 neighborhoods with the most reported crime events among all 296 neighborhoods of LA. On average, there are $4,140$ crime events in the top $100$ neighborhood and $8,750$ such events in the top $10$. We then apply the community-detection methods mentioned in \cref{sec:community} to the reconstructed networks; this assigns crime topics to communities. We quantify the difference between these community assignments and the crime-topic classifications from \cite{kuang_crime_2017} by calculating NMI. We also visualize the crime-topic networks of the Westwood and Wingfoot neighborhoods in \cref{fig:westwood}; they are, respectively, located in West LA and South LA. From \cref{tab:nmi2}, we see that using spatial information combined with a nonparametric kernel leads to the best mean NMI score among the methods that we examine.

\begin{table}
	\centering
	\footnotesize{
	\caption{Mean NMI (with one standard deviation reported in parentheses) between community assignments from several community-detection methods and the classifications from \cite{kuang_crime_2017} in the 100 neighborhoods in Los Angeles with the most recorded crime events between 1 January 2009 and 19 July 2014. 
			\label{tab:nmi2}}
			\begin{tabular}{|c|c|c|c|}
				\hline
				&  Nonparametric  &  Temporal & Parametric \\
				\hline
				Symmetric NMF & 0.25 (0.11) & 0.12 (0.084) &    0.084 (0.12)\\ 
				\hline
				Weighted SBM  &  0.24 (0.12) & 0.085 (0.086) &   0.078 (0.079) \\
				\hline
		\end{tabular}
	}
\end{table}


\subsection{Network of Crime Events} \label{sec:NCE}

In the previous sections, we studied relationships among entities given spatiotemporal events associated with them. To examine connections between events, we now define an event network, which is both weighted and directed, in which each event is a node and $\mathbf{P}$ denotes the adjacency matrix of this network. Recall that, in \cref{subsubsection: pm-em}, we constructed an expectation matrix $\mathbf{P}$, where $\mathbf{P}(i,j)=p_{ij}$ is the probability that event $j$ is triggered by event $i$ and $\mathbf{P}(i,i)=p_{ii}$ is the probability that event $i$ is a background event. The weight of an edge reflects a triggering effect between two events, and the direction points from the earlier event to the later one. For example, we can build a crime-event network in which each node is a crime incident (i.e., an event), and we estimate edges between events using our nonparametric model.


\subsubsection{Stochastic Declustering}\label{declust}

With an event network, a natural question is whether one can differentiate between ``true'' background events and triggered events. Such differentiation using the probability $p_{ii}$ is called \textit{stochastic declustering} \cite{zhuang2002stochastic}. To determine whether event $i$ is a background event, we compare $p_{ii}$ with a uniformly random sample from the interval $(0,1)$. If $p_{ii}$ is larger than the random number, we consider this event to be from the background; otherwise, we consider it to be triggered by other events.

We perform declustering experiments on synthetic data; we simulate ten synthetic point processes using a fixed triggering matrix that we generate from a WSBM. (See \cref{subsec:syn} for details.) Recall from \cref{alg:simu} that we retain causality information in the simulations (i.e., which events cause which others and which events are from the background), giving a notion of ``ground truth'' about the ancestors of each event. One way to measure the quality of declustering is by comparing the inferred branching ratio \cite{sornette2009limits} with the one from the ground-truth data. The branching ratio is defined as $1-{N_b}/{N}$, where $N_b$ is the number of background events. However, the difference in branching ratios itself typically does not completely reflect reconstruction errors. For example, in an extreme case, stochastic declustering can erroneously misclassify some number of background events as triggered and the same number of triggered events erroneously as background, although the branching ratio is the same as the true branching ratio in this scenario. To resolve this problem, we view declustering as a binary classification problem that assigns events to be either background or triggered events. We use measurements such as recall and precision to evaluate our declustering results. Recall that ``recall'' is the ratio between the number of background events that are correctly recovered by the declustering methods (i.e., the true positives) to the total number of background events, and ``precision'' is the ratio between the number of true positives to the number of events that are labeled as background events by stochastic declustering. From the results in \cref{tab:decl}, we see that the Temporal Hawkes method has the worst performance among the methods that we consider. Our Nonparametric Hawkes method has the best recall and precision (with the smallest variations as well), and the Parametric Hawkes method has the smallest branching-ratio error.

\begin{table}
	\centering
	\footnotesize{
		\caption{Comparison of our stochastic declustering results for the Nonparametric Hawkes, Parametric Hawkes, and Temporal Hawkes methods using synthetic point-process data with networks from a WSBM (see \cref{subsec:syn}) and background labels from the simulation from \cref{alg:simu}. We report the mean and the standard deviation (in parentheses) of the branching-ratio error, precision, and recall over ten simulations (which we do for ten point processes with the same triggering kernels and matrix). For each simulation, each calculation is the mean over $20$ runs of stochastic declustering. 
			\label{tab:decl}}
		
		\begin{tabular}{|c|c|c|c|}   
			\hline
			& Nonparametric &  Parametric & Temporal \\
			
			\hline
			Branching-ratio error&     0.039 (0.0050)
			 &     0.01 (0.011)
			&    0.022 (0.019)
			\\ 
			\hline
			Recall       
			&   0.75 (0.0098)
			&0.65 (0.027)
			&    0.60 (0.035)
			\\ 
			\hline
			Precision
			&   0.70 (0.0082)
			&    
			0.64 (0.0093)
			&    0.59 (0.0086)
			\\ 
			\hline
	\end{tabular}}
\end{table}


 \subsubsection{Motif Analysis}

Declustering methods can help differentiate between background and triggered events in an event network. To further examine spatiotemporal dynamics, we consider causality information among events. Similar to a relational-event model \cite{butts2008relational}, one can obtain causality information from the matrix $\textbf{P}$, because $p_{ij}$ is the probability that event $j$ is triggered by event $i$. We focus on repeated patterns to obtain information about local causality structure. Specifically, we examine network motifs \cite{milo2002network}, which are recurrent (and often statistically significant) patterns in a network. Note that all event networks are directed acyclic graphs (DAGs) because of how they incorporate temporal information.

We find that motif analysis is insightful for studying gang-crime event networks. Gang crimes are often characterized by retaliations (triggered crime events) among rivalry gangs; this can lead to a series of tit-for-tat reciprocal crimes. To find significant gang retaliation patterns, we use a gang-crime data set (provided by the LAPD) from 2014--2015 with 4,158 events in Los Angeles. Using these data, we generate an event network with our Nonparametric Hawkes method. We then threshold the network, by keeping edges whose weight is at least $0.1$ and then binarizing them, so that the edges are unweighted. We use the motif-detection method and code \cite{mfinder} from \cite{milo2002network}, including their null model.\footnote{For each of our networks, we produce $100$ ``randomized'' networks. To produce one such network, we use the default edge-swapping approach from \cite{milo2002network}. This entails making a number of random swaps equal to about 100--200 times the number of edges. For each node in a network, we require that the randomized network preserves its number of in-edges, out-edges, and bidirectional edges.} 

We find, for thresholds ranging from $0.5$ to $0.001$, that a three-node feedforward-loop motif \cite{mangan2003structure} occurs more significantly than by chance (with z-scores that are larger than $2$) in both the city-wide data set and in the South LA\footnote{South LA Gang Reduction and Youth Development (GRYD) Zones\cite{GRYD17}.} subset (which consists of 1,912 events) of the data set. Davies and Marchione~\cite{davies2015event} found that the same three-node motif is significant in networks that they constructed (using different methods from ours for both network construction and motif detection) using data sets from maritime piracy and residential burglaries.

\begin{figure}
		\centering 
        \label{fig:3motifs}
		\includegraphics[width=0.8\linewidth]{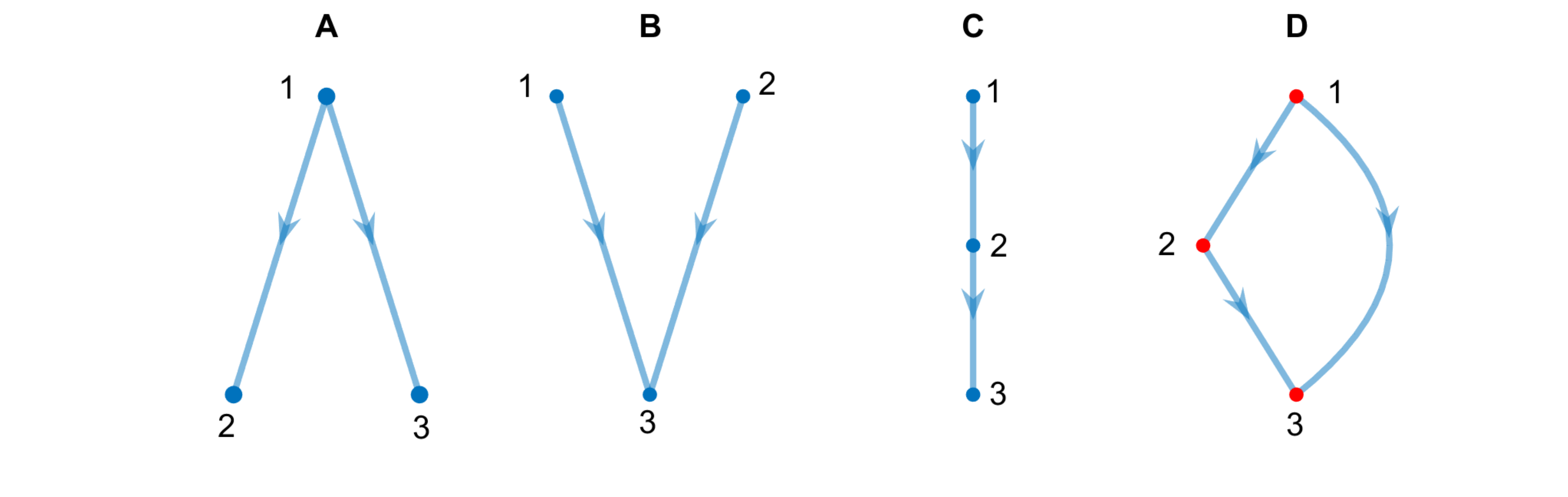}
	\caption{All possible three-node motifs for a DAG. We highlight the nodes in the feedforward-loop motif (D) in red.
	} 
\end{figure}

We focus on the South LA area, because it is the center of a gang intervention program \cite{GRYD17}. Establishing which causal structures are statistically significant has important implications for countering gang violence, and fast response to gang crime may reduce the potential that it triggers a future retaliation. Knowing that feedforward-loop network motifs occur at rates that are larger than chance suggests that disrupting retaliation may require assessment of trade-offs in how to allocate intervention resources. For example, in a simple triggering chain (see \cref{fig:3motifs}C), one can expect that intervention following an initial triggering event in time will have a direct effect on the second event in time and an indirect effect on the third event, although the effect on the third event may be attenuated by the intervening event. By contrast, we expect that intervention following the first event in the feedforward-loop motif (see \cref{fig:3motifs}D) will have a direct effect on the second event in time and both a direct and indirect effect on the third event. It is possible that the third event is more likely to be disrupted given the feedforward structure and intervention following the first event than would be the case with direct intervention following only the second event.


\section{Conclusions and Discussion} \label{sec:conclusions}

In this paper, we used point-process models to infer latent networks from synthetic and real-world spatiotemporal data sets. We then applied tools from network analysis to examine the inferred networks. We studied the role of spatial information and nonparametric techniques in network reconstruction.

As we have illustrated, it is very important to incorporate spatial information. However, using such information effectively requires making a good choice of spatiotemporal triggering kernels. We achieved this using a nonparametric approach. Through experiments on synthetic data sets, we showed that our nonparametric Hawkes method is capable of doing a good job of successfully recovering spatial and temporal triggering kernels. Moreover, our approach is able to infer a network structure that better recovers --- compared to other network reconstruction methods that we studied --- symmetry and reciprocity, edge reconstruction, and community structures. Through experiments on real-world data sets, we illustrated that the inferred networks of our approach are meaningful, in the sense that they have large positive correlations with some metadata. 

This paper helps fill a gap on incorporating spatial information into multivariate self-exciting point processes \cite{reinhart2018rejoinder}, and it will be interesting to apply our approach to other fields (such as seismology). Moreover, our approach is not limited to the Euclidean distance (for the spatial variables) that is used commonly in seismology\cite{ogata_space-time_1998} and crime applications\cite{mohler_self-exciting_2011}. In other words, although the spatial triggering kernel $g_2(r)$ in this paper is a function of Euclidean distance, one can potentially use any notion of ``distance'' between two entities. For example, in a network, one can measure a distance between two entities based on the length of shortest paths between them. In a recent paper, Green et al. \cite{green2017modeling} proposed a social-contagion model in which they assumed, using a parametric form, that the strength of triggering in a Hawkes-process model depends on the shortest-path distance. With our approach, we can nonparametrically estimate such dependence. To give another example, consider a point process in which each event is associated with textual information. For instance, in a Twitter data set, one can consider each tweet (a time-stamped body of text) as an event in a point process. One can measure a distance between two tweets based on their text. 

Naturally, our network reconstruction method is not without limitations. It uses $O(U^2)$ parameters for $U$ nodes. To avoid underfitting, it requires a large number of observed events. The computational complexity and memory requirement scale at least quadratically with the number of events, so the current EM-type algorithm is not ideal for analyzing large data sets. It will thus be important to improve our inference method for network reconstruction.


\appendix

\section{Preprocessing of the Gowalla Data} \label{appA}

\begin{figure}
	\centering
	\subfloat[1-Ego network of a user of Gowalla in SF.]{\includegraphics[width=0.3\textwidth]{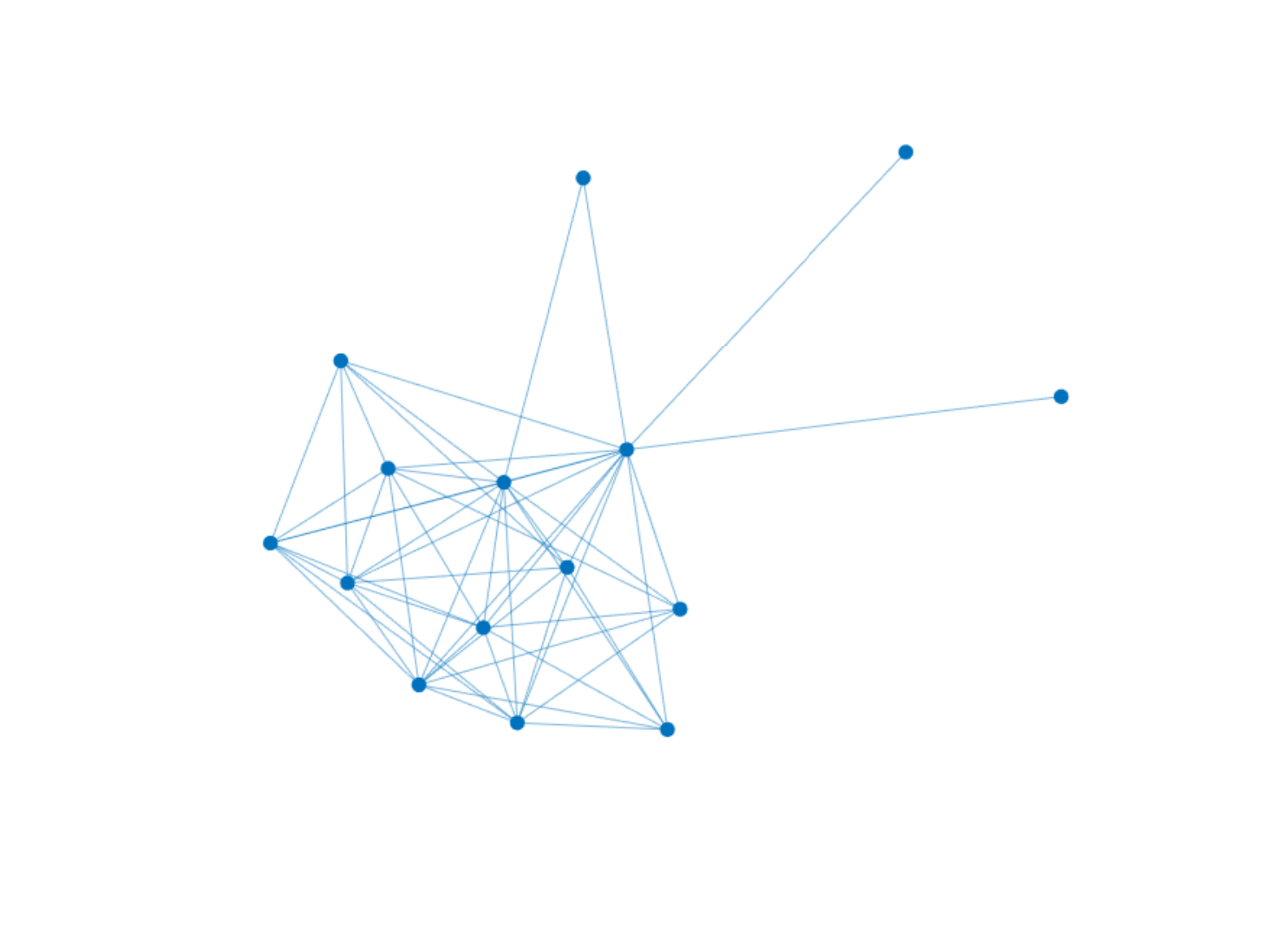}}\hspace{2mm}
	\subfloat[Largest connected component of the Gowalla network in NYC.]{\includegraphics[width=0.3\textwidth]{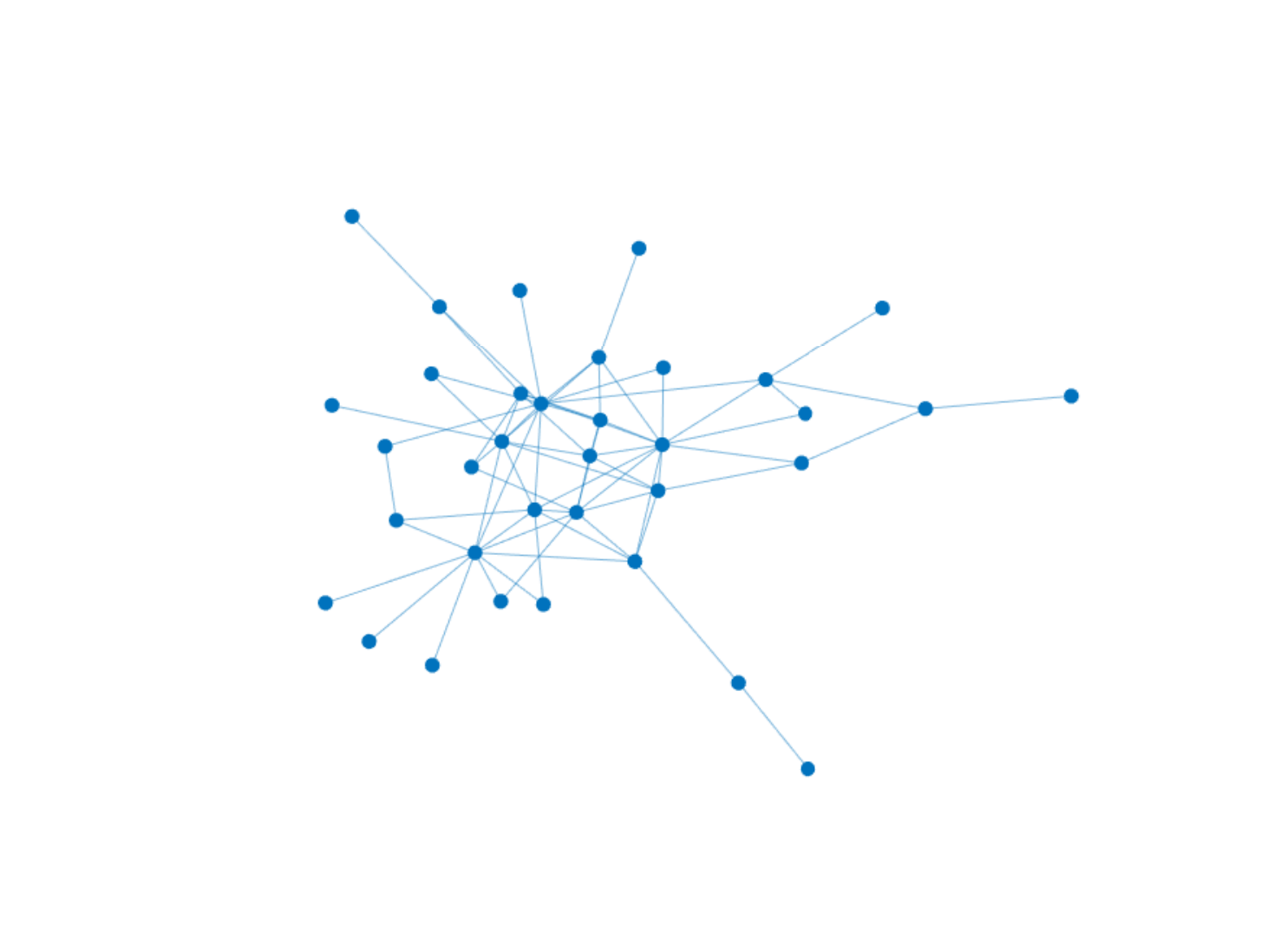}}\hspace{2mm}
	\subfloat[Largest connected component of the Gowalla network in LA.]{\includegraphics[width=0.3\textwidth]{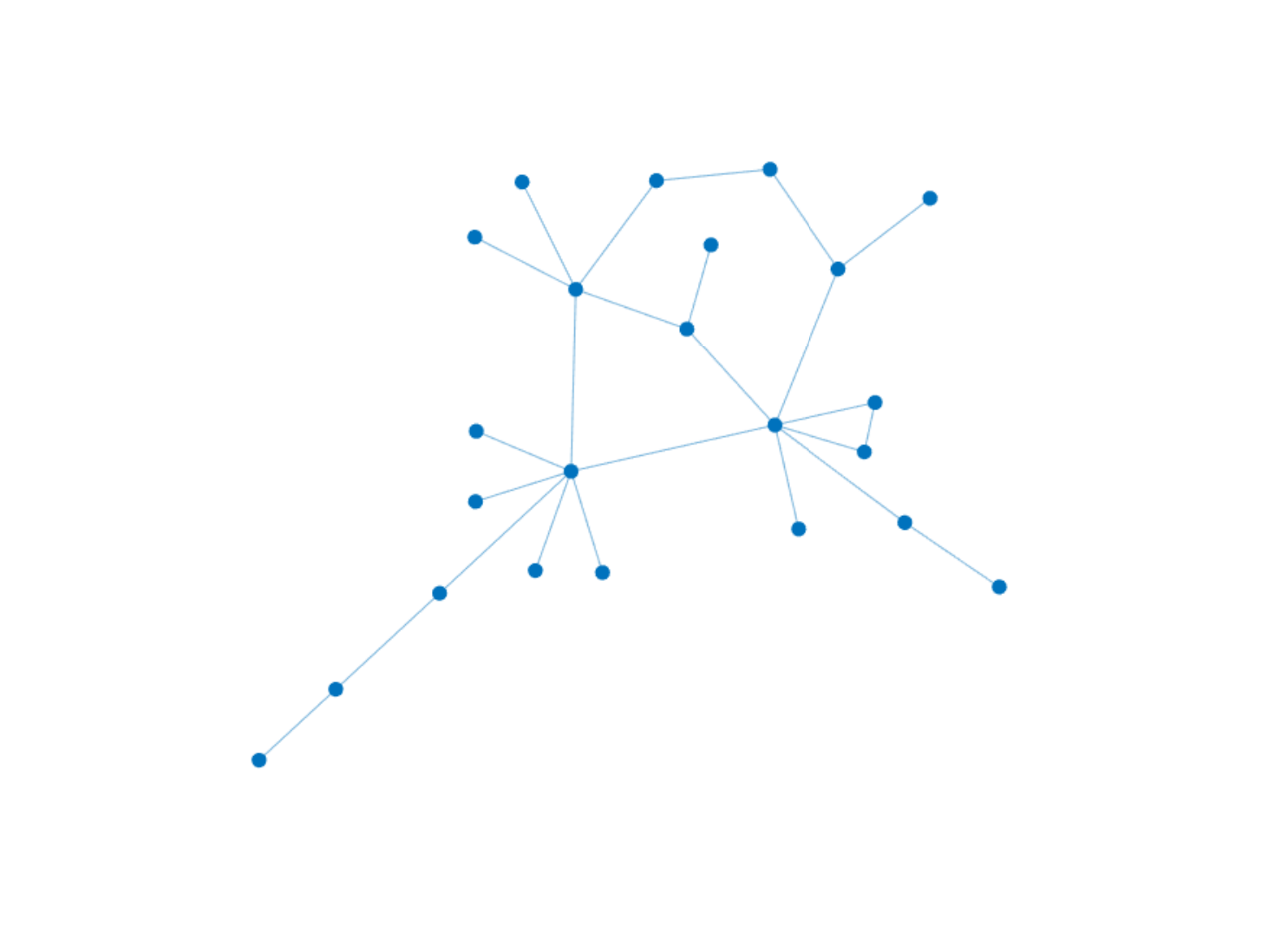}}
	\caption{Three different friendships networks in the Gowalla data set. We compare different network reconstruction methods for these networks.  
	}
	\label{fig:ego}
\end{figure}

In this section, we detail how we preprocess the Gowalla data that were collected and studied in \cite{cho_friendship_2011}. We examine data from three cities: New York City, Los Angeles, and San Fransisco. We visualize the networks used in this paper in \cref{fig:ego}.


\subsection{New York City (NYC)} \label{nyc}

We study check-ins in New York City (NYC) during the period April--October 2010. We use a bounding box (with a north latitude of $40.92$, a south latitude of $40.48$, an east longitude of $-73.70$, and a west longitude of $-74.26$)\footnote{We obtain latitude and longitude coordinates from \protect\url{http://www.mapdevelopers.com/geocode_bounding_box.php}.} to locate check-ins in NYC. We consider ``active'' users, who have at least $100$ check-ins during the period. To alleviate the computational burden, we also only consider users who have at most $500$ check-ins during the period to reduce the number of users and the total number of check-ins. Our inference process requires computing a triggering probability for each pair of events (i.e., check-ins), which results in a full upper-triangular matrix. The number of nonzero entries in this matrix scales with the square of the total number of events, so the memory requirement also scales quadratically with the number of events. We perform experiments only for cases in which the total number of events is at most $10,000$ to be able to store triggering probabilities for all pairs of events in 4-gigabyte memory. There are $5,801$ unique users with at least one check-in in NYC during the period, and there are $101,329 check-ins$ in total. After removing ``inactive'' users (i.e., those with strictly fewer than $100$ check-ins) and overly active users (i.e., those with strictly more than $500$ check-ins), we are left with $160$ users and a total of $29,118$ check-ins. We also restrict ourselves to users in the largest connected component (LCC) of the network. This yields $46$ users and $8,495$ check-ins, on which we apply our inference methodology.  


\subsection{Los Angeles (LA)}

We apply the same procedure as in \cref{nyc} on the check-in data for Los Angeles (LA). The bounding box that we use for LA has a north latitude of $34.34$, a south latitude of $33.70$, an east longitude of $-188.16$, and a west longitude of $-188.67$. We restrict the area of LA to be the same as that of NYC, although LA's geographic area is much larger than that of NYC. After selecting only users in the LCC of the Gowalla network among users who are active (with at least $150$ check-ins) but not overly active (with at most $1000$ check-ins) users, we are left with $23$ users and $6,203$ check-ins.


\subsection{San Francisco (SF)} 

To look at a different type of example, we also examine the 1-ego network of the most popular user (with 14 friends) in San Francisco (SF). (A 1-ego network \cite{ugander2011anatomy} of a node is an induced subgraph that includes a focal node --- the ego --- and its direct neighbors.) 
The bounding box that we use for SF has a north latitude of $37.93$, a south latitude of $37.64$, an east longitude of $-122.28$, and a west longitude of $-123.17$. In this 1-ego network, there are $9,887$ check-ins.

\section*{Acknowledgements}

We thank Chandan Dhal  and Jialin Liu for helpful discussions and preliminary work on community detection.





\end{document}